\documentclass[conference]{IEEEtran}
\IEEEoverridecommandlockouts
\usepackage{cite}
\usepackage{amsmath,amssymb,amsfonts}
\usepackage{algorithmic}
\usepackage{graphicx}
\usepackage{textcomp}
\usepackage{xcolor}
\usepackage{comment}
\usepackage{tabularx}
\usepackage{multirow}
\usepackage{array}
\usepackage{float}
\usepackage{placeins}
\usepackage{balance}
\usepackage{stfloats}
\usepackage{graphbox}
\usepackage{mdframed}
\usepackage{colortbl}
\usepackage{url}

\def\BibTeX{{\rm B\kern-.05em{\sc i\kern-.025em b}\kern-.08em
    T\kern-.1667em\lower.7ex\hbox{E}\kern-.125emX}}
\begin{document}

\vspace{-2cm}
\title{Code Reviewer Recommendation Based on a Hypergraph with Multiplex Relationships 
}

\author{
    \IEEEauthorblockN{Yu Qiao$^{1}$, Jian Wang$^{1*}$,
    Can Cheng$^{2*}$, Wei Tang$^{1}$, Peng Liang$^{1}$, Yuqi Zhao$^{1}$, Bing Li$^{1*}$}
    \IEEEauthorblockA{$^1$ School of Computer Science, Wuhan University, Wuhan, China}
    \IEEEauthorblockA{$^2$ School of Artificial Intelligence, Hubei University, Wuhan, China}
    \IEEEauthorblockA{\{qiaoyu\_cs, jianwang, weitang\_cs, liangp, yuqizhao, bingli\}@whu.edu.cn, cancheng@hubu.edu.cn}
}


\maketitle

\begin{abstract}
Code review is an essential component of software development, playing a vital role in ensuring a comprehensive check of code changes. However, the continuous influx of pull requests and the limited pool of available reviewer candidates pose a significant challenge to the review process, making the task of assigning suitable reviewers to each review request increasingly difficult. To tackle this issue, we present MIRRec, a novel code reviewer recommendation method that leverages a hypergraph with multiplex relationships. MIRRec encodes high-order correlations that go beyond traditional pairwise connections using degree-free hyperedges among pull requests and developers. This way, it can capture high-order implicit connectivity and identify potential reviewers. To validate the effectiveness of MIRRec, we conducted experiments using a dataset comprising 48,374 pull requests from ten popular open-source software projects hosted on GitHub. The experiment results demonstrate that MIRRec, especially without PR-Review Commenters relationship, outperforms existing state-of-the-art code reviewer recommendation methods in terms of ACC and MRR, highlighting its significance in improving the code review process.

\end{abstract}


\section{Introduction}
Modern Code Review (MCR) \cite{b1} refers to examining source code systematically to identify and correct overlooked mistakes. This practice is indispensable in the pursuit of enhancing software quality. MCR is typically carried out before code integration and release, varying in granularity from individual lines of code to entire software modules. During this process, one or more reviewers assess the code for errors, adherence to coding standards, test coverage, and more. The proliferation of the pull-request (PR) mechanism has reshaped the landscape of massive open-source projects, establishing PR as a critical tool for the implementation of MCR. The shift has made MCR more collaborative and continuous \cite{b2}. Developers now have the ability to create a PR and request reviewers who possess the expertise and relevant past review experience to inspect the uploaded code changes. This PR-centric review mechanism helps improve code quality, ensure code standards, promote team collaboration, and accelerate the development process.

Despite its advantages, the PR-centric review mechanism can also be challenging. It requires developers to dedicate time and effort that could be otherwise spent writing new code. In some cases, reviewers invited to review the code may decline due to factors like unfamiliarity with the content or lack of time. This presents a human-related issue in matching PRs to appropriate reviewers, which can be labor-intensive and time-consuming\cite{b3}. Consequently, the task of identifying suitable reviewers, while recognized as crucial within the OSS community, remains a significant challenge.

\textbf{Limited reviewer capacity}. Reviews on PRs enable collaborators to comment on, approve, or request further modifications to the proposed changes before merging. However, this workflow, while effective, increases the workload for reviewers. The rapid growth of PRs, especially in large and popular projects, coupled with the need for collaboration among several reviewers for PRs that affect multiple modules, poses a significant challenge for timely responses. For instance, Tencent, one of the largest Chinese Internet companies, receives over $100$K code changes per month in proprietary projects \cite{b4}. Moreover, reviewer teams often find themselves significantly outnumbered by the sheer volume of incoming PRs. A case in point is React, a popular open-source project on GitHub, which has received over $12$K PRs since its inception in March $2023$, with fewer than $1$K developers participating in the review process. Yang et al. \cite{b2} also reported that within three years, $437$ reviewers contributed more than 66K reviews to the LibreOffice project. 
These examples highlight the challenges faced by reviewer teams in managing and responding to PRs efficiently.

\textbf{High review latency}. Previous studies\cite{b3}\cite{b5}\cite{b6}\cite{b7} have shown that effective reviews require knowledgeable or familiar reviewers for the submitted changes. Inappropriate reviewers can hinder the review process, delay PR merges, and slow down development. Rigby and Bird\cite{b8}  observed that 50\% of reviews take nearly 30 days. Similarly, Tsay et al.\cite{b9} found that a number of code changes remain pending for up to two months before being merged. Additionally, reviewers occasionally decline review requests\cite{b10} , further exacerbating review delays. Another study indicated that 16\%-66\% of the patches have at least one invited reviewer who did not respond to the review invitation\cite{b3}. The decision of whether a reviewer accepts a PR review invitation depends on various factors, including their available time, existing workload, review experience, code authoring experience, and more \cite{b3}. Hence, recommending appropriate reviewers for incoming PRs is considered a practical and collaborative way to expedite meaningful merges and facilitate the OSS evolution.

In response to these issues, researchers have conducted numerous studies. Appropriate reviewers who are more familiar with the code changes may spend less time reviewing them\cite{b11}\cite{b12}\cite{b13}\cite{b14}. However, Thongtanunam et al. \cite{b12} found that about 4\%-30\% of reviews encounter code-reviewer assignment problems, and larger projects face greater difficulties in finding suitable code reviewers. Existing reviewer recommendation approaches focused on modeling the developer’s expertise or interactions with tasks based on historical information \cite{b4} \cite{b15} \cite{b16} \cite{b17}. However, these methods heavily rely on historical data to model a developer's expertise and task interactions. If the historical data is insufficient or inaccurate, it may negatively impact the quality of recommendations. Different developers may also possess distinct expertise and patterns of task interactions. If a model cannot effectively capture these individual differences, it may affect the accuracy of the recommendations. 

\begin{figure}[http]
\vspace{-0.3cm}
\centerline{\includegraphics[width=6cm,height=2cm]{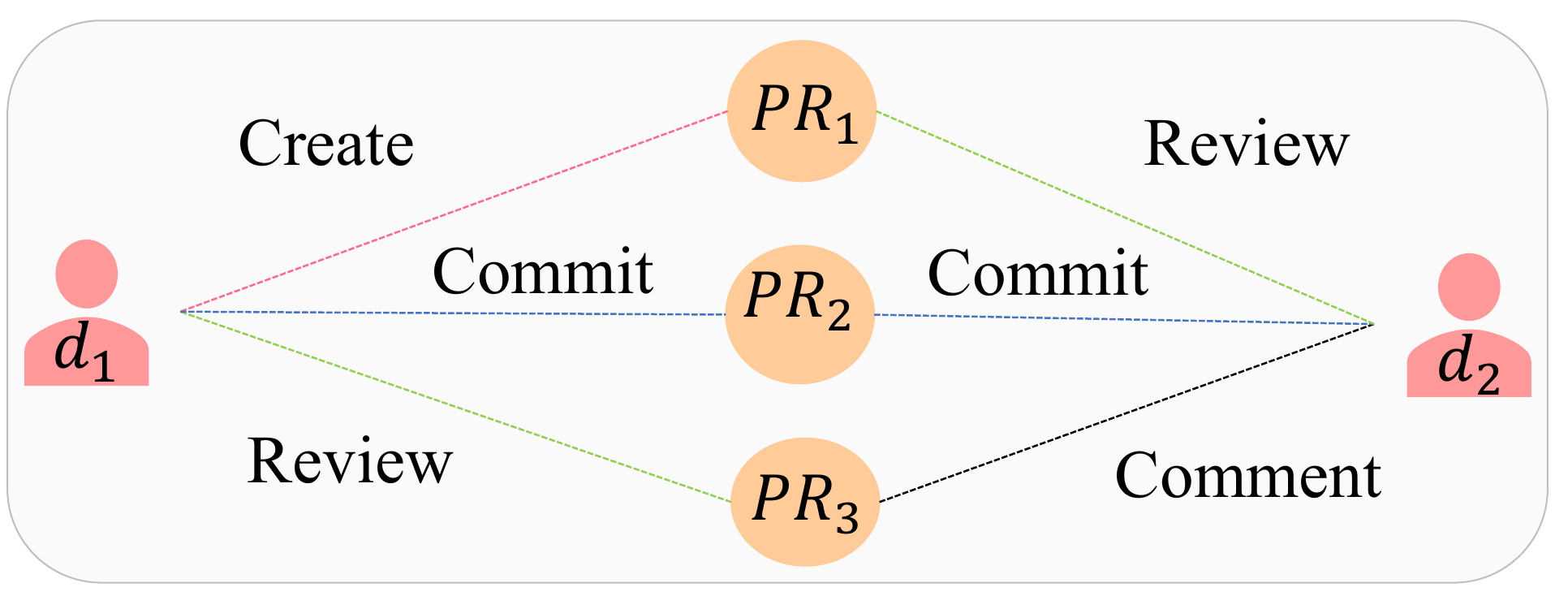}}
\vspace{-0.3cm}
\caption{Multiplex interactions of developers in PR-based review process} 
\vspace{-0.2cm}
\label{fig1}
\end{figure}

Furthermore, recommendation models based on relation graphs or networks~\cite{b18} \cite{b19}\cite{b20}\cite{b21}\cite{b22} proposed to alleviate the sparse explicit interactions by considering various explicit and implicit relationships. 
Specifically, Yu et al. \cite{b18} constructed a social relation network among PR, creators, and reviewers, while Rong et al. \cite{b22} developed a hypergraph reviewer recommender based on three types of relations.
Although these studies have made important contributions to reviewer recommendation, there is one notable limitation that much of the current research primarily treats interactions equally, without considering the varying roles that developers play.  
Actually, developers may transition between these roles across different PRs. For instance, developer $d_1$ might act as a reviewer for $PR_3$, while being a committer for $PR_2$, as depicted in Figure~\ref{fig1}. 
These multifaceted roles and behaviors of developers, which can represent their abilities in various development tasks, are often overlooked in existing code reviewer recommendation methods. 
Another limitation is that these existing recommendation approaches based on relation networks often combine traditional methods, such as Collaborative Filtering and Graph Neural Networks, to extract features from a bipartite graph. However, given the inherent challenges in comprehending developer expertise, relevance, and the sparsity of explicit developer-task interactions, these approaches are struggling to comprehensively capture these relations and thus encounter difficulties in achieving high performance.

To this end, this work aims to automate the recommendation of potentially qualified reviewers for PRs. To address the issues aforementioned, we apply the hypergraph to model the multivariate interactions among developers and PRs, enabling us to recommend appropriate reviewers for PRs. To evaluate the effectiveness of the proposed approach, we conducted several experiments on ten OSS projects and compared it with the state-of-the-art methods. The main \textbf{contributions} of this paper include:

$(1)$ We propose MIRRec, a hypergraph method accomplished by considering the PR similarity and the \underline{M}ultiplex \underline{I}nteraction \underline{R}elationships among developers and PRs within the various roles that developers play, to \underline{Rec}ommend potential reviewers for PRs.

$(2)$ We validate the effectiveness of MIRRec on ten open-source projects, and the results indicate that MIRRec outperforms the state-of-the-art approaches.


$(3)$ Investigating the influence of interactions on MIRRec reveals that the type and overlap of interaction data can influence reviewer recommendation performance, providing valuable insight for the development of further models.

The rest of the paper is organized as follows. Section \ref{section2} delves into the details of our method. Section \ref{section3} outlines the experiment design of evaluation. Section \ref{section4} scrutinizes the outcomes of our evaluation. Section \ref{section5} highlights the potential threats to validity. In Section  \ref{section6}, we discuss prior research and key distinctions in our approach. Finally, Section \ref{section7}  reports conclusions and puts forward future research.

\section{Approach}\label{section2}

\subsection{Approach overview}\label{subsec3.A}
\begin{figure*}[htbp]
\vspace{-0.6cm}
\centerline{\includegraphics[width=12.5cm,height=6.5cm]{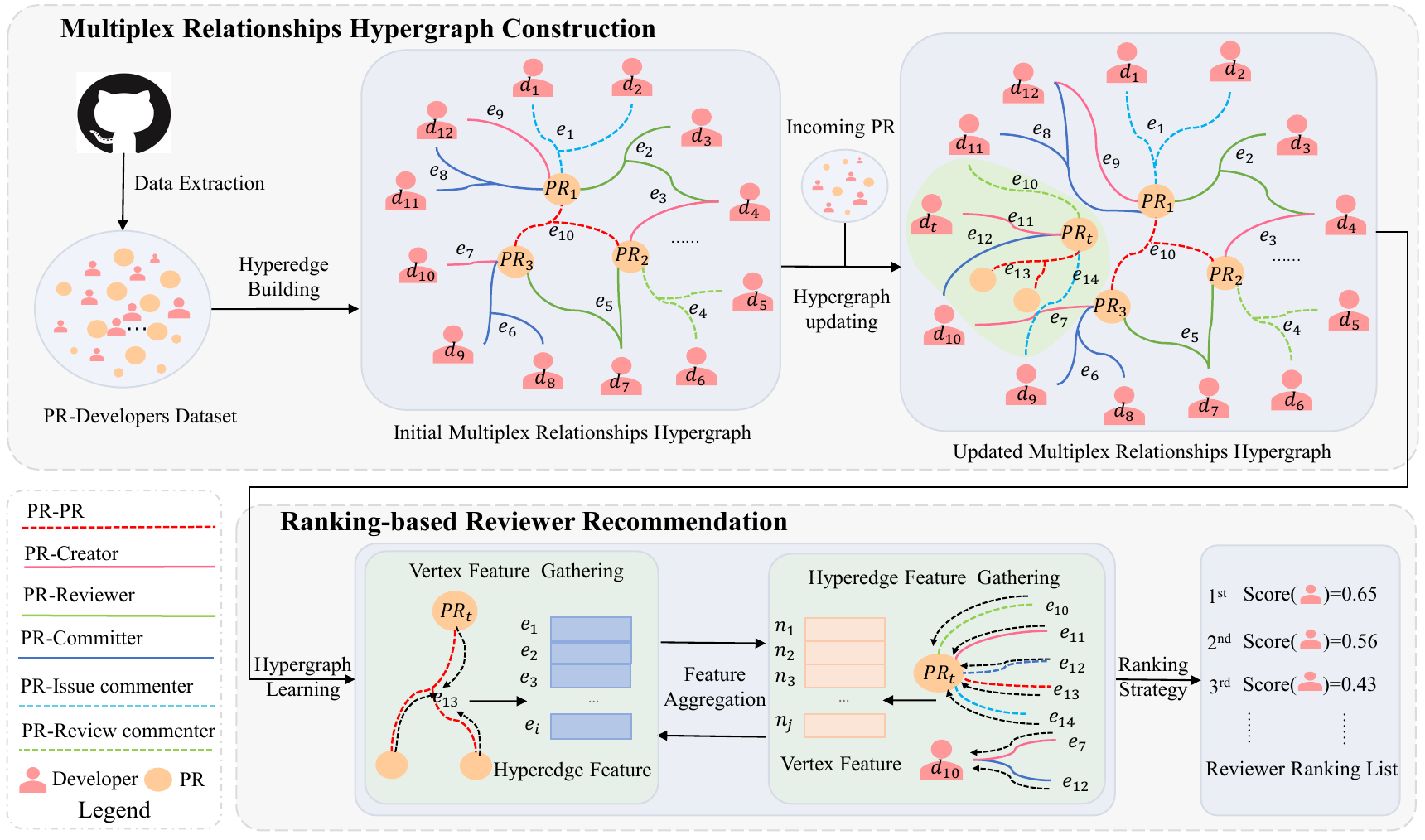}}
\vspace{-0.3cm}
\caption{Overview of approach} 
\vspace{-0.6cm}
\label{fig2}
\end{figure*}

As depicted in Figure \ref{fig2}, our proposed MIRRec comprises two key steps: multiplex-relationship hypergraph construction and ranking-based reviewer recommendation. We begin by identifying PRs and developers as graph nodes, creating hyperedges among them and assigning different weights based on their interaction history and task similarity to establish the multiplex-relationship hypergraph. Next, we update the existing hypergraph with incoming PR data and employ a hypergraph-based learning strategy to calculate ranking scores efficiently for ranking and recommending suitable reviewers for the new PR. 

\subsection{Multiplex-relationship Hypergraph Construction }
Developers mainly participate in the pull and review processes of PRs through create, commit, comment, and review, while playing different roles as creator, committer, commenter, and reviewer~\cite{b23}. Consequently, we build the following six hyperedges, inspired by these behaviors associated with the different roles developers assume, as well as task (PRs) similarity, and assign the weights respectively.

\textbf{PR-Creator hyperedge}: We establish a hyperedge between each PR and its creator. Each PR is created by a single developer, who is also one of the committers for the PR and is familiar with the files and code modifications involved in the PR. The creation behavior implies the developer’s preference and the scope of development tasks in the current time point more accurately. Therefore, we determine the weight of each PR based on the time elapsed since its creation relative to the start of the dataset. More recent creations receive higher weights. The weight is calculated as:
\vspace{-0.2cm}
\begin{align}\label{eq1}
\resizebox{.3\hsize}{!}{
$W(p_{i}, u_{j}) = \frac{t_{i} - t_{s}}{t_{e} - t_{s}}$},
\end{align}
where $t_s$ and $t_e $ represent the start time and end time of the dataset, while $t_i$ is the creation time of PR $p_i$ by creator $u_j$.

\textbf{PR-Committers hyperedge}: A PR typically includes several commits, which may be made by different developers. These committers are very likely to participate in the review of the relevant file and codes of the PR as they are more familiar with this PR. Therefore, we establish a hyperedge connecting the PR with its associated committers. We calculate the weight of the hyperedge considering the modified code lines, the commit creation time, and the number of commits\cite{b24}, as formulated in the following equation:
\begin{align}\label{eq2}
\resizebox{.9\hsize}{!}{
$W(p_{i},u)_{ct} = \sum_{m = 1}^{N_{(p_{i},u)}^{ct}}{\sum_{n = 1}^{N_{(p_{i},u_{m})}^{ct}}{\alpha^{n - 1} \times \frac{\left| t_{mn}^{ct} - t_{s} \right|}{t_{e} - t_{s}}}} \times \frac{1}{1 + e^{0.01 \times l_{code}}}$},
\end{align}
where $N_{(p_{i},u)}^{ct}$ is the total count of committers for $p_i$. $N_{(p_{i},u_{m})}^{ct}$ is the commit count by committer $u_m$ to $p_i$. $t_{mn}^{ct}$and $l_{code}$ are the $n$-th commit time and the modification code lines of $u_m$. $\alpha\in[0,1]$ is an experience attenuation coefficient used to adjust the influence of committers with multiple commits on the hyperedge.

\textbf{PR-Reviewers hyperedge}: PRs often undergo several rounds of revisions and code reviews before being merged. This is likely accomplished through the cooperation of different reviewers. These review interactions are essential for reviewer recommendation tasks. Here, we build a PR-Reviewers hyperedge between the PR and its associated reviewers. The weight of a PR-Reviewer edge is set as follows:
\begin{align}\label{eq3}
\resizebox{.9\hsize}{!}{
$W(p_{i},u)_r = \sum_{m = 1}^{N_{(p_{i},u)}^r}{\sum_{n = 1}^{N_{(p_{i},u_{m})}^{r}}\alpha^{n - 1}} \times e^{\frac{t_{mn}^r - t_{s}}{t_{e} - t_{s}} - 1}$}, 
\end{align}
where $N_{(p_{i},u)}^r$ is the reviewers number of $p_i$. $N_{(p_{i},u_{m})}^{r}$ denotes the review times conducted by $u_m$ for $p_i$. $t_{mn}^r$ is the start time of $n$-th review by $u_m$ on $p_i$.

\textbf{PR-Issue Commenters hyperedge and PR-Review Commenters hyperedge}: PR is a kind of issue with code or file modifications. Developers use comments to suggest improvements, ask questions, or provide feedback on these changes in the pull and review processes. These comments help improve code quality and facilitate collaboration and communication among project members. Additionally, some of these commenters may transition into code reviewers. There are two types of comments on PRs: comments directly on the PR, which can also be considered as issue comments, and comments specifically on code or file modifications within the PR review, which serve as review comments. Therefore, we establish these two hyperedges through connections between the PR and its respective commenters. We use the creation time and the number of comments as the basis for measuring the interest of developers, and define the formulas for calculating the hyperedge weights of PR-Issue Commenters and PR-Review Commenters. as follows:
\begin{align}\label{eq4}
\resizebox{.9\hsize}{!}{
$W(p_{i},u)_{ic} = \sum_{m = 1}^{N_{(p_{i},u)}^{ic}}{\sum_{n = 1}^{N_{(p_{i},u_{m})}^{ic}}\alpha^{n - 1}} \times e^{\frac{t_{mn}^{ic} - t_{s}}{t_{e} - t_{s}} - 1}$}, 
\end{align}
where $W(p_{i},u)_{ic}$ represents the weight of PR-Issue Commenters hyperedge. $N_{(p_i,u)}^{ic}$ is the count of commenters of $p_i$. $N_{(p_i,u_m)}^{ic}$ is the number of comments by $u_m$ on $p_i$. $t_{mn}$ is the comment time of $n$-th comment by $u_m$ on $p_i$. Similarly, the weight of PR-Issue Commenters hyperedge $W(p_{i}, u)_{rc}$ is calculated in the same way.

It is important to note that our hypotheses for \eqref{eq2}, \eqref{eq3}, and \eqref{eq4}  are as follows: $(1)$ Interactions (commits, reviews, and comments) that developers engage in across multiple PRs hold greater significance compared to interactions within a single PR. $(2)$ Recent interactions carry more weight than older ones. To distinguish the variations in interactions between multiple PRs and those within a single PR, we introduce an empirical decay factor $\alpha$ to control their influence. Additionally, we incorporate the time-sensitive factor (the partition similar as \eqref{eq1}) to prioritize recent interactions\cite{b18}\cite{b22}.

\textbf{PR-PR hyperedge}: Developers are likely to review PRs that are similar to previously reviewed PRs, as they possess the necessary expertise. Here, we create PR-PR hyperedges connecting similar PRs, with the weight determined by the similarity of their modified file path sets\cite{b12}, which is:
\vspace{-0.2cm}
\begin{align}\label{eq5}
\small
& W(p_{i},p_{j})  = \\
& \begin{aligned}
\begin{cases}
0 &  F_{i}=0 | F_{j}=0 \nonumber\\
\sum\limits_{f_{m}\in F_{i}}\sum\limits_{f_{n}\in F_{j}}\frac{\text{SimFilePath}(f_{m},f_{n})}{|F_{i}||F_{j}|}\times e^{-\frac{|t_{i}-t_{j}|}{t_{e}-t_{s}}} &  F_{i}\geq 0| F_{j}\geq 0,
\end{cases}
\end{aligned}
\end{align}

\vspace{-0.5cm}

\begin{align}\label{eq6}
\small
\text{SimFilePath}(f_{m},f_{n}) = \frac{\text{LCP}(f_{m},f_{n})}{\max(\text{len}(f_{m}),\text{len}(f_{n}))},
\end{align}
where $F_i$ and $F_j$ represent the modified file path set of $p_i$ and $p_j$ respectively. $t_i$ and $t_j$ are the creation time of  $p_i$ and $p_j$. $\text{SimFilePath}(f_m,f_n)$ is the distance between file $f_m$ and $f_n$. $\text{LCP}(f_m,f_n )$ denotes the longest common prefix function.

As the similarity between PRs is affected by the modified file path set and creation time, the similarity between most PRs in the dataset is low. Thus, we select Top-K similar PRs for each PR to build a PR-PR hyperedge (cf. following for details) and preferentially consider the latest PRs.

\subsection{Ranking-based Reviewer Recommendation}

We carry out the review recommendation as a ranking task on the constructed hypergraph. The key is to find an optimal ranking vector $f^{*}\in \mathbb{R}^{V}$ to rank the matching scores between reviewers and PRs. The details are elaborated as follows.

\subsubsection{Hypergraph learning}
For the hypergraph we constructed $G = \langle V,E,W\rangle$, which includes a vertex set $V$, a hyperedge set $E$, and each hyperedge is assigned a weight by $W$. The key of ranking is to minimize the object function as below:
\setlength{\abovedisplayskip}{1pt}
\setlength{\belowdisplayskip}{1pt}
\begin{align}\label{eq7}
\psi(f) = \varphi(f) + R_{\text{emp}}(f),
\end{align}
where $\psi(f)$ is a regularization on hypergraph, $R_{\text{emp}}(f)$ denotes the supervised empirical loss. $\varphi(f)$ is defined as:
\begin{align}\label{eq8}
\small
\varphi(f) &= f^{T}(I-D_{v}^{-1/2}HWD_{e}^{-1}H^{T}D_{v}^{-1/2})f \nonumber\\
&= f^{T}(I-A)f \nonumber\\
&= f^{T}Lf
\end{align}
where $H$ represents the incidence matrix between $V$ and $E$, $D_{v}$ and $D_{e}$ denote the vertex degree matrix and hyperedge degree matrix respectively, and $A$ is the adjacency matrix of $G$.

To recommend reviewers for a newly incoming PR $p_i$, the ranking loss function is represented as:
\begin{align}\label{eq9}
\small
R_{\text{emp}}(f) = \lambda||f-y_{p_{i}}||^{2},
\end{align}
$\lambda$ is the positive parameter to weigh the supervised empirical loss.
Then the object function $(f)$ can be deducted to:
\begin{align}\label{eq10}
\psi(f) = f^{T}Lf + \lambda||f-y_{p_{i}}||^{2}
\end{align}

The optimal $f^*$  can be deducted and transformed into:
\begin{equation}\label{eq11}
f^{*} = (I-\mu A)^{-1}y_{p_{i}},
\end{equation}
where $\mu\in[0,1]$ is the hyperparameter and $\mu=1/(1+\lambda)$.

\subsubsection{Ranking strategy}

Having ranked on the hypergraph, we can recommend the Top-K reviewers as the candidates. We define the ranking score by considering the relevance of reviewer candidates in different roles excluding creators to the PR:
\begin{equation}\label{eq12}
\text{Score}_{u_{i}} = af^{*}[r_{i}] + bf^{*}[ct_{i}] + cf^{*}[rc_{i}] + df^{*}[ic_{i}],
\end{equation}
where $f^{*}[r_{i}]$,$f^{*}[ct_{i}]$,$f^{*}[rc_{i}]$,$f^{*}[ic_{i}]$ represent the relation scores as reviewers, committers, review commenters and issue commenters of $u_i$ respectively. If a developer has had no interaction with a PR in a specific role mentioned above, the corresponding relation score is set to $0$. The variables $a$, $b$, $c$, and $d$ are the weights assigned to each relation. The Top-K code reviewers are then recommended based on the ranking scores of the code reviewer candidates.

\section{Evaluation}\label{section3}
This study aims to propose a relationship-aware and hypergraph-based approach for code reviewer recommendation and assess the impact of the relationship-based hypergraph strategy on recommendation performance. To this end, three research questions (RQs) are defined to evaluate MIRRec.

\textbf{RQ1: Which hyperparameter settings can enhance the performance of MIRRec?}

\textit{Rationale}: Identifying optimal hyperparameter settings helps improve the performance of MIRRec. With this RQ, we want to use efficient methods to determine these settings within limited computational resources and investigate the model’s behavior influenced by the hyperparameters.

\textbf{RQ2: Does MIRRec surpass state-of-the-art reviewer recommendation methods in terms of performance?}

\textit{Rationale}: Different from other state-of-the-art methods, MIRRec incorporates multiplex relationships and hypergraph learning. With this RQ, we aim to evaluate the effectiveness of this strategy.

\textbf{RQ3: How do different types of interaction relationships affect the recommendation performance of MIRRec?}

\textit{Rationale}: As found by Ruangwan et al. in~\cite{b3}, different relationships may have varying impacts on reviewer recommendation. Therefore, this RQ aims to understand the impact of interaction relationships on the recommendation performance.

RQ1 aims to identify hyperparameter settings that can enhance MIRRec's performance. RQ2 aims to validate the effectiveness of the optimal hyperparameter settings found in RQ1 by comparing them with baseline methods. RQ3 further explores how different types of relationships influence MIRRec's performance, thus providing additional support for the results in RQ2 using the optimal settings obtained in RQ1.

\subsection{Data Preparation}

We focus on popular and active OSS projects, as they tend to have frequent code updates and a wealth of PR and developer data. For this purpose, we selected OSS projects from GitHub for dataset based on the following criteria: $(1)$ projects created before Oct $1$st, $2017$ with last commit posted after Aug $1$st, $2022$, indicating a sufficiently long lifespan and continuous updates; $(2)$ projects with more than $10$K stars and $3$K forks, indicating high popularity; and $(3)$ projects with over $3$K PRs and 8K review histories, ensuring sufficient data for analysis.

Considering factors like programming language popularity, star and fork rankings, and the quantity of PR-review interaction histories, we selected a final set of ten OSS projects spanning six programming languages. To avoid PRs with early data imbalance (a small number of PRs in the early stages of project creation) and recent data instability (some PRs not yet closed), we used data from June $1$st, $2018$, to Nov $30$th, $2021$, for this study.

\subsection{Data Preprocessing}
We performed several preprocessing steps on the source data to construct the dataset, following these steps:

$\bullet$ \textbf{Unifying developer identities.} Developers have four identity attributes: `ID', `Login', `Name', and `Email'. `ID', `Login', and `Email' serve as unique identifiers for developers. However, the `Email' attribute may be empty, and the `Name' attribute may be updated or duplicated. Additionally, different developer roles may have incomplete identity attributes. For instance, committers only have `Name' and `Email' attributes. These variations make it challenging to accurately map the same developers who play different roles and exhibit different interaction behaviors based solely on their identity attributes. To address this issue, we employed the following steps for unique identifier mapping: 

$(1)$ For `Name' attributes containing `and', `\&', `\&\&', `+', or `|', we utilized regular expressions to split them into separate data elements. $(2)$ We created a mapping table that connects the `Login', `Email', and `Name' attributes of developers involved in the project. This table enabled us to perform conversions effectively. In most cases, developers can obtain their unique `Login' based on the mapping table for a given project. $(3)$ If a developer cannot be successfully mapped using the previous steps, we calculated the edit distance \cite{b25} for `Login', `Email', and `Name' attributes. We chose the attribute with a shorter edit distance for mutual conversion. $(4)$ Despite the above processing steps, there may still be approximately 1\% of developers who cannot be identified through the mapping. For these developers, we considered them as other independent developers.

$\bullet$ \textbf{Filtering out invalid interaction data.}
We identify that some developers are invalid, and their interaction histories can interfere with the recommendation task. To address this, we removed the invalid interaction data related to three types of developers: $(1)$ Robot developers: In large OSS projects, there are robots performing code testing and reviewing. We removed data associated with developers whose type is listed as `Bot' in the developer information.  $(2)$ Deregistered developers: Historical review records of logged-out users can interfere with the recommendation. GitHub's API marks the status of logged-out users as empty. Therefore, we deleted any data where the developer information is empty. $(3)$ Developers who reviewed their own PRs: In most cases, PR creators were directly involved in code changes. If creators also reviewed their own PRs, their subjectivity could significantly impact the quality of code reviews. Hence, we discarded code review interaction data where the reviewer and the creator are the same.

$\bullet$ \textbf{Removing commit data with bulk file modifications.}
When developers make modifications to source code, they may end up modifying a significant number of files due to changes in widely used classes or functions, as well as folder renaming. While many of these modifications may be insignificant, they can result in multiple interactions with these files by committers, potentially impacting the effectiveness of the reviewer recommendation. To mitigate this, we excluded commit interaction histories where the number of modified files equals or exceeds $100$.

$\bullet$ \textbf{Discarding data after PRs are merged.} After PRs are merged, it is common for developers to continue leaving comments. However, since the merging of PRs signifies the end of the review phase, the comments generated after the merge become less relevant for code reviewer recommendations. As a result, we discarded any relevant data that occurs after the PR has been merged.

$\bullet$ \textbf{Removing PRs without file modifications or review history.}
To construct the hypergraph, it is crucial to consider the value of the interaction data. Some PRs have not undergone a review process, lack actual code changes, or do not involve file submissions. These types of PRs include non-code changes (such as documentation, comments, or README files), branch merges (integrating changes between branches without explicit file modifications), and collaborative discussions (primarily for discussions and reviews without direct code modifications or file submissions). The relationships captured in these types of PRs are often distant and less informative. Therefore, we excluded these PRs and their associated data.

After applying the aforementioned preprocessing steps, we obtained the dataset, and its details are presented in Table \ref{table1}.

\begin{table}[htbp] 
\vspace{-0.5cm}
  \centering
  \caption{Overview of the dataset}
  \vspace{-0.2cm}
  \fontsize{7pt}{10pt}\selectfont
  \setlength{\tabcolsep}{1pt} 
  \begin{tabularx}{\linewidth}{X*{8}{>{\centering\arraybackslash}X}} 
    \hline
    Language & Fork & Star & Project & PR & Commit & Issue \mbox{comment} & Review \mbox{comment} & Review \\
    \hline
    \multirow{4}{*}{C++} & 34.3 k & 69.7 k & Bitcoin & 4544 & 11053 & 26587 & 22106 & 20088 \\
         & 14.7 k & 108 k  & Electron & 10163 & 23058 & 12053 & 15526 & 21050 \\
         & 55.3 k & 69 k   & Opencv   & 3158 & 8452  & 5050  & 9886  & 8205 \\
         & 6.2 k  & 15.8 k & XBMC     & 2789 & 5422  & 7378  & 8324  & 7760 \\
    \hline
    \multirow{2}{*}{JavaScript} & 43.4 k & 208 k & React   & 3167 & 8371  & 4661  & 6693  & 6773 \\
               & 23.6 k & 88.3 k & Angular & 9095 & 17370 & 14737 & 29288 & 25883 \\
    \hline
    Python & 29.2 k & 71 k & Django & 2861 & 4302 & 5892 & 11258 & 8262 \\
    PHP    & 9.1 k  & 28.4 k & Symfony & 7420 & 9814  & 16084 & 21689 & 25142 \\
    Ruby   & 21.1 k & 52.9 k & Rails  & 2812 & 4861  & 5029  & 7708  & 6793 \\
    Scala  & 3.2 k  & 14.1 k & Scala  & 2365 & 4537  & 5398  & 5429  & 5866 \\
    \hline
    \multicolumn{4}{c}{total} & 48374 & 97240 & 102869 & 137907 & 135822 \\
    \hline
  \end{tabularx}
  \label{table1}
  \vspace{-0.5cm}
\end{table}

\subsection{Experiment settings}
We notice that different projects can exhibit varying numbers of PRs and levels of developer activity over time. For instance, the Scala project received approximately $500$ PRs within a year, whereas the Electron project received a similar number within just two months. Developers may also go through periods of extended inactivity within a project. In the case of the Django project, we identify a total of $1039$ developers, with the majority showing short-term activity. However, only 71 developers displayed a pattern of short-term activity followed by a prolonged period of inactivity (one year) before reengaging with the project. To account for these variations, we divided the data into training and testing sets on an annual basis. For example, the training set encompasses the period from June $2018$ to June $2019$, while the testing set consists of data from July $2019$. This strategy enabled us to generate $30$ rounds of sliding training and testing data, with each round representing a monthly interval, and calculate the average performance as the final evaluation result.

To evaluate MIRRec, we compared this method with several state-of-the-art code reviewer recommenders, which include:

$(1)$ \textbf{RevFinder} \cite{b12}, which recommends code reviewers based on the similarity of the file path sets of PR modifications. This approach aims to recommend code reviewers who have expertise in handling PRs with similar file paths in historical review records.

$(2)$ \textbf{cHRev} \cite{b15}, which scores candidate reviewers based on their specific contributions like expertise, workload, and frequency of past reviews.

$(3)$ \textbf{CN} \cite{b18}, which recommends developers with similar interests as reviewers by constructing a social relation network based on the interaction between PR creators and reviewers in the review history. 

$(4)$ \textbf{HGRec} \cite{b22}, which recommends reviewers using a hypergraph-based approach that considers the interactions of contributors and reviewers with PR, as well as the similarity between PRs.

We select these baselines for the following reasons: RevFinder and cHRev have frequently served as the basis for comparison in many existing studies. CN adopts a graph as the underlying model, while HGRec is the first to use a hypergraph in the reviewer recommendation task. For these baseline methods, we either made minimal modifications to their source code to accommodate our dataset or reproduced based on their research paper. We use Accuracy (ACC) and Mean Reciprocal Rank (MRR), which are commonly used \cite{b26}, to evaluate the performance of MIRRec. 
The replication package of this study has been publicly available \cite{b27}.

\section{Results}\label{section4}
\subsection{Results for RQ1: Hyperparameters Study}
\begin{figure}[http]
\vspace{-0.4cm}
\centerline{\includegraphics[width=8.5cm,height=3.5cm]{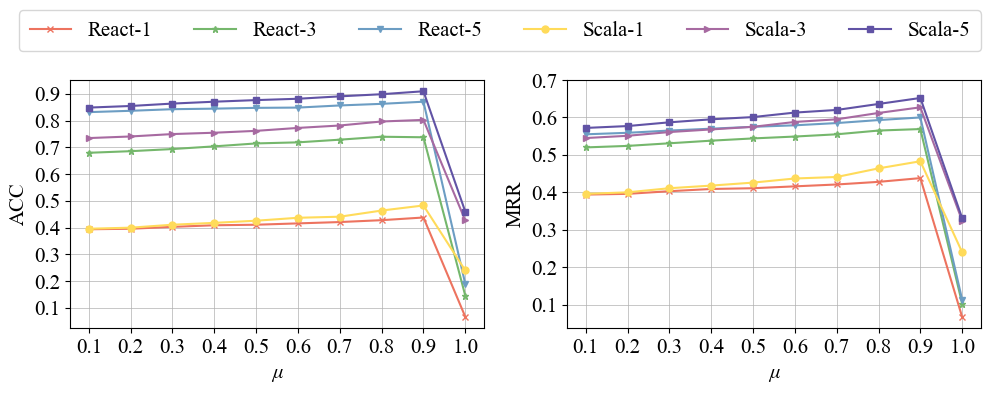}}
\vspace{-0.4cm}
\caption{ ACC and MRR performance of MIRRec under different $\mu$ } 
\vspace{-0.3cm}
\label{fig3}
\end{figure}

Due to limited computational resources, we employ a greedy search strategy to configure the parameters $\mu$, K, and $\alpha$. As for the parameters $a$, $b$, $c$, and $d$, which play a role in the ranking calculation after graph learning, we utilize a grid search strategy to determine the optimal parameter combinations. The specific details are outlined below:

$\bullet$  The hyperparameter $\mu \in [0,1]$ is used to adjust the impact of the standardized loss function and the empirical loss function. Due to the high experimental cost of seeking the optimal solution within a continuous value range, we conduct experimental analysis on discrete values in steps of $0.1$  within the value range on different projects. Different values of $\mu$ have varying impacts on the performance of MIRRec across different projects. However, MIRRec consistently exhibits the same trend across various projects, achieving optimal performance when $\mu$ is $0.9$. Figure \ref{fig3} shows the change curves of the Top-$1$\texttildelow Top-$5$ and MRR of MIRRec on the projects of Scala and React under different hyperparameter $\mu$ values\footnote{Due to space limitations, we randomly selected several projects to demonstrate the impact of various hyperparameters on the model}. Therefore, \textbf{we set the value of hyperparameter$\mu$ to 0.9 for subsequent experiments}. 

$\bullet$ The parameter K represents the maximum number of connections a PR can have in the PR-PR hyperedge. Having too few connections can result in inadequate learning of higher-order relationships between similar PRs and developers, while too many connections can interfere with recommendations and increase model training time. MIRRec aims to recommend a list of reviewers (Top-$1$\texttildelow Top-$5$) for PRs based on multiple interactions between PRs and developers. To obtain higher-order relationships involving at least five reviewers through at least five similar PRs, we set the hyperparameter K between $5$ and $25$ using a ‘trial-and-error’ approach to achieve optimal recommendation performance.

\begin{figure}[http]
\vspace{-0.3cm}
\centerline{\includegraphics[width=8.5cm,height=4cm]{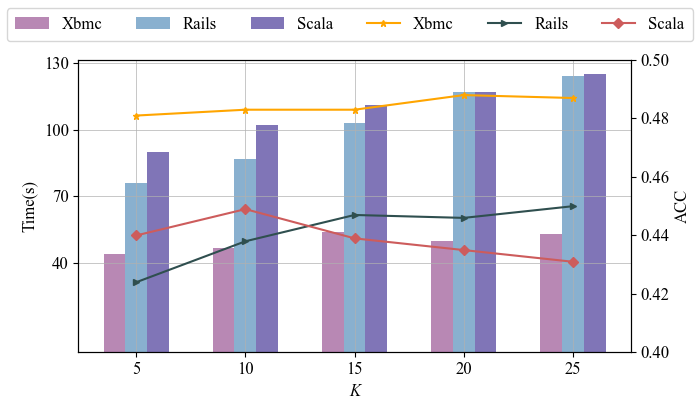}}
\vspace{-0.3cm}
\caption{Training time and ACC performance of MIRRec under different K} 
\label{fig4}
\vspace{-0.3cm}
\end{figure}

As shown in Figure \ref{fig4}, the line data and column data represent the model accuracy (Top-$1$) and training time under different K values, respectively. Within the range of $5$ to $10$, increasing K leads to improvements in both accuracy and training time. However, when K exceeds $10$, the model's accuracy on different projects fluctuates, with either a slight improvement or significant decrease, while the training time continues to rise. \textbf{Thus, we ultimately determined 10 as the optimal value for hyperparameter K and use it for subsequent experiments}.

$\bullet$ The parameter $\alpha$  represents the influence a reviewer poses in a history review. To understand how different $\alpha$ settings affect the performance of MIRRec, this article also adopted a ‘trial-and-error’ method. We observe the ACC for different projects at Top-$1$ to Top-$5$, and the coefficient of variation\cite{b28} for ACC is consistently below $0.01$, as shown in Table \ref{table2}. \textbf{This indicates that $\alpha$ has a limited overall impact on the performance of MIRRec in different projects}. Further literature research reveals that many reviewer recommendation methods based on interaction relationships use an empirical value of $0.8$ for $\alpha$ \cite{b22}. \textbf{Therefore, we set $\alpha$ to $0.8$ in our subsequent experiments}.


\begin{table}[htbp]
  \centering
   \vspace{-0.45cm}
  \caption{coefficient of variation for ACC}
  \vspace{-0.2cm}
    \begin{tabular}{cccccc}
    \hline
    Project & Django & React & XBMC & Scala & Rails \\
    \hline
    Top-1 & 0.002390 & 0.004174 & 0.003025 & 0.009531 & 0.013636 \\
    Top-3 & 0.001733 & 0.003600 & 0.001658 & 0.003270 & 0.003438 \\
    Top-5 & 0.000753 & 0.002401 & 0.001199 & 0.000858 & 0.003369 \\
    \hline
    \end{tabular}%
  \label{table2}%
  \vspace{-0.15cm}
\end{table}%

$\bullet$ 
To explore the impact of four interaction relationships that is, PR-Reviewers, PR-Committers, PR-Review Commenters, and PR-Issue Commenters, on reviewer recommendation, we conducted multiple experiments on the entire dataset based on interaction relationship hypergraphs with different weight combinations ($a$, $b$, $c$, $d$ in \eqref{eq12}) to optimize performance. Statistical analysis of the dataset reveals that over $75\%$ of developers who served as reviewers also served as committers. In comparison, only $25\%$ of developers who acted as issue commenters have served as reviewers. Additionally, a substantial overlap was observed between reviewers and creators, as well as between committers and reviewers. This suggests that the PR-Reviewers and PR-Committers relationships exert a more significant influence on the reviewer recommendation compared to the other types of comment-related relationships.

Since there was no obvious way to obtain the optimal weight combination, we conducted a pilot study through `trial and error' experiments. We found that the overall performance was better when c:d$\approx$1:1, Thus, we further conducted `trial and error' experiments with the premise of $c$:$d$=$1$:$1$, \textbf{and finally found that $a$: $b$: $c$: $d$=$4$:$3$:$1$:$1$ resulted in better model performance}. The partial results are shown in Table~\ref{table3}.

\begin{table}[tp]
\vspace{-0.5cm}
  \centering
  \vspace{-0.15cm}
  \caption{MIRRec Performance under different weight combinations(\%)}
  \vspace{-0.2cm}
  \fontsize{8pt}{9pt}\selectfont
  \renewcommand{\arraystretch}{1}
    \begin{tabularx}{\linewidth}{*{4}{>{\centering\arraybackslash}p{0.2cm}}*{6}{>{\centering\arraybackslash}X}}
    \hline
    a & b & c & d & $ACC_1$ & $ACC_3$& $ACC_5$ & $MRR_1$ & $MRR_3$  & $MRR_5$\\
    \hline
    1  & 1  & 1  & 1  & 46.5 & 80.3 & 90.4 & 46.5 & 61.7 & 64 \\
    2  & 1  & 1  & 1  & 46.7 & 80.4 & 90.6 & 46.7 & 61.9 & 64.3 \\
    3  & 1  & 1  & 1  & 47.2 & 80.3 & 90.8 & 47.2 & 62.2 & 64.6 \\
    4  & 1  & 1  & 1  & 47.1 & 79.7 & 90.7 & 47.1 & 62.0  & 64.5 \\
    5  & 1  & 1  & 1  & 47.4 & 80.2 & 90.8 & 47.4 & 62.3 & 64.7 \\
    1  & 2  & 1  & 1  & 46.2 & 80.1 & 90.4 & 46.2 & 61.4 & 63.8 \\
    2  & 2  & 1  & 1  & 47.1 & 79.9 & 90.6 & 47.1 & 62.1 & 64.6 \\
    3  & 2  & 1  & 1  & 48.0  & 80.0  & 90.7 & 48.0  & 62.5 & 65.0 \\
    4  & 2  & 1  & 1  & 47.6 & 80.0  & 90.6 & 47.6 & 62.4 & 64.8 \\
    5  & 2  & 1  & 1  & 47.6 & 80.1 & 90.9 & 47.6 & 62.4 & 64.9 \\
    1  & 3  & 1  & 1  & 45.0  & \textbf{80.9} & 89.7 & 45.0  & 60.9 & 63.0 \\
    2  & 3  & 1  & 1  & 46.8 & 80.4 & 90.6 & 46.8 & 61.9 & 64.3 \\
    3  & 3  & 1  & 1  & 48.1 & 80.2 & 90.2 & 48.1 & 62.6 & 65 \\
    4  & 3  & 1  & 1  & \textbf{48.3} & 80.3 & \textbf{91} & \textbf{48.3} & \textbf{62.7} & \textbf{65.2} \\
    5  & 3  & 1  & 1  & 48.1 & 79.9 & 90.8 & 48.1 & 62.5 & 65.0 \\
    \hline
  \end{tabularx}%
  \label{table3}%
  \vspace{-0.4cm}
\end{table}%

\begin{mdframed}[backgroundcolor=gray!20]
\textbf{Answer}: The hyperparameter $\mu$, K, $\alpha$ were set to $0.9$, $10$, $0.8$, and  $a$: $b$: $c$: $d$ was set to $4$:$3$:$1$:$1$  for MIRRec determined by 'trial and error’ experiments, demonstrating consistent and superior performance across projects.
\end{mdframed}

\subsection{Results for RQ2: Overall Performance Comparison}

\begin{table*}[hbp]
\vspace{-0.6cm}
  \centering
  \caption{Comparison of ACC across Different Methods (\%)}
  \vspace{-0.2cm}
  \renewcommand{\arraystretch}{1}
    \begin{tabularx}{\linewidth}{X*{15}{>{\centering\arraybackslash}X}} 
    \hline
    \multirow{2}{*}{Project} & \multicolumn{3}{c}{RevFinder} & \multicolumn{3}{c}{cHRev} & \multicolumn{3}{c}{CN} & \multicolumn{3}{c}{HGRec} & \multicolumn{3}{c}{MIRRec} \\
\cline{2-16}
 & Top-1 & Top-3 & Top-5 & Top-1 & Top-3 & Top-5 & Top-1 & Top-3 & Top-5 & Top-1 & Top-3 & Top-5 & Top-1 & Top-3 & Top-5 \\
    \hline
    Bitcoin & 31.1 & 58.1 & 69.7 & 24.5 & 50.7 & 65.7 & \cellcolor{gray!15}32.0  & \cellcolor{gray!15}58.2 & \cellcolor{gray!15}71.3 & \cellcolor{gray!35}34.0  & \cellcolor{gray!35}61.8 & \cellcolor{gray!35}73.1 & \cellcolor{gray!60}\textbf{34.8} & \cellcolor{gray!60}\textbf{63.2} & \cellcolor{gray!60}\textbf{74.8} \\
    
    Electron & \cellcolor{gray!15}36.4 & 75.1 & 90.5 & 16.3 & 51.4 & 78.1 & \cellcolor{gray!60}\textbf{41.7} & \cellcolor{gray!15}76.5 & \cellcolor{gray!35}92.8 & \cellcolor{gray!60}\textbf{41.7} & \cellcolor{gray!35}77.1 & \cellcolor{gray!15}92.7 & \cellcolor{gray!35}41.5 & \cellcolor{gray!60}\textbf{78} & \cellcolor{gray!60}\textbf{93.1} \\
    
    Opencv & 58.9 & 83.5 & 90.9 & 47.7 & 75.6 & 84.1 & \cellcolor{gray!60}\textbf{69.4} & \cellcolor{gray!15}88.1 & \cellcolor{gray!15}92.8 & \cellcolor{gray!15}65.3 & \cellcolor{gray!35}88.8 & \cellcolor{gray!35}94.8 & \cellcolor{gray!35}68.8 & \cellcolor{gray!60}\textbf{89.8} & \cellcolor{gray!60}\textbf{95.5} \\
    
    XBMC & 29.0  & 55.5 & 67.6 & 34.0  & 59.3 & 71.5 & \cellcolor{gray!15}42.8 & \cellcolor{gray!15}63.5 & \cellcolor{gray!15}72.9 & \cellcolor{gray!35}43.2 & \cellcolor{gray!35}66.6 & \cellcolor{gray!35}76.8 & \cellcolor{gray!60}\textbf{44.9} & \cellcolor{gray!60}\textbf{68.7} & \cellcolor{gray!60}\textbf{78.7} \\
    
    React & 24.3 & 56.0  & 74.9 & 22.2 & 54.8 & 71.2 & \cellcolor{gray!35}43.6 & \cellcolor{gray!60}73.9 & \cellcolor{gray!15}85.5 & \cellcolor{gray!15}41.7 & \cellcolor{gray!15}73.2 & \cellcolor{gray!35}86.1 & \cellcolor{gray!60}\textbf{43.8} & \cellcolor{gray!35}\textbf{73.8} & \cellcolor{gray!60}\textbf{87.1} \\
    
    Angular & 29.6 & 61.6 & \cellcolor{gray!15}77.6 & 32.6 & 63.5 & 77.4 & \cellcolor{gray!15}40.2 & \cellcolor{gray!15}66.5 & 76.5 & \cellcolor{gray!35}46.8 & \cellcolor{gray!35}72.2 & \cellcolor{gray!35}83.4 & \cellcolor{gray!60}\textbf{48.5} & \cellcolor{gray!60}\textbf{73.8} & \cellcolor{gray!60}\textbf{84.1} \\
    
    Django & 46.9 & \cellcolor{gray!15}79.3 & \cellcolor{gray!15}87.6 & 36.0  & 69.5 & 81.6 & \cellcolor{gray!15}57.0  & 77.8 & 85.1 & \cellcolor{gray!35}59.1 & \cellcolor{gray!35}81.7 & \cellcolor{gray!35}87.9 & \cellcolor{gray!60}\textbf{59.9} & \cellcolor{gray!60}\textbf{81.9} & \cellcolor{gray!60}\textbf{88.1} \\
    
    Symfony & \cellcolor{gray!15}46.8 & \cellcolor{gray!15}79.3 & \cellcolor{gray!15}87.1 & 36.8 & 68.3 & 80.5 & 45.4 & 77.2 & 85.7 & \cellcolor{gray!35}48.2 & \cellcolor{gray!35}80.8 & \cellcolor{gray!35}89.1 & \cellcolor{gray!60}\textbf{49.5} & \cellcolor{gray!60}\textbf{82.2} & \cellcolor{gray!60}\textbf{89.5} \\
    
    Rails & \cellcolor{gray!60}\textbf{24.7} & \cellcolor{gray!35}45.2 & \cellcolor{gray!35}58.9 & 14.7 & 34.4 & 47.6 & 20.9 & 41.7 & 54.5 & \cellcolor{gray!15}21.9 & \cellcolor{gray!15}44.9 & \cellcolor{gray!15}58.6 & \cellcolor{gray!35}22.3 & \cellcolor{gray!60}\textbf{47.4} & \cellcolor{gray!60}\textbf{59.6} \\
    
    Scala & 43.1 & 74.2 & 86.5 & 28.8 & 63.3 & 77.7 & \cellcolor{gray!15}45.5 & \cellcolor{gray!15}77.5 & \cellcolor{gray!15}87.4 & \cellcolor{gray!35}46.9 & \cellcolor{gray!35}79.3 & \cellcolor{gray!35}90.3 & \cellcolor{gray!60}\textbf{48.3} & \cellcolor{gray!60}\textbf{80.3} & \cellcolor{gray!60}\textbf{91.0} \\
    \hline
    Average & 37.1 & 66.8 & 79.1 & 29.4 & 59.1 & 73.5 & \cellcolor{gray!15}43.9 & \cellcolor{gray!15}70.1 & \cellcolor{gray!15}80.5 & \cellcolor{gray!35}44.9 & \cellcolor{gray!35}72.6 & \cellcolor{gray!35}83.3 & \cellcolor{gray!60}\textbf{46.2} & \cellcolor{gray!60}\textbf{73.9} & \cellcolor{gray!60}\textbf{84.2} \\
    \hline
    Improve & 24.53\% & 10.63\% & 6.45\% & 57.14\% & 25.04\% & 14.56\% & 5.24\% & 5.42\% & 4.60\% & 2.90\% & 1.79\% & 1.08\% & -  &  -  & - \\
    \hline
  \multicolumn{16}{p{0.95\linewidth}}{%
  \raggedright
  \textbf{Note}:`Average' indicates the mean performance across the entire dataset.
  `Improve' signifies the degree to which MIRRec outperforms other methods in terms of average performance.
  Bold indicates the optimal result, and three gray background colors, from light to dark, are used to highlight the top three results (Third, Second, First) in terms of performance for Top-1, Top-3, and Top-5, respectively.
  } \\
  \hline
   \end{tabularx}
  \label{table4}
  \vspace{-0.3cm}
\end{table*}

\begin{table*}[ht]
\vspace{-0.6cm}
  \centering
    \caption{Comparison of MRR across Different Methods (\%)}
  \vspace{-0.2cm}
  \renewcommand{\arraystretch}{1}
    \begin{tabularx}{\linewidth}{X*{15}{>{\centering\arraybackslash}X}} 
    \hline
    \multirow{2}{*}{Project} & \multicolumn{3}{c}{RevFinder} & \multicolumn{3}{c}{cHRev} & \multicolumn{3}{c}{CN} & \multicolumn{3}{c}{HGRec} & \multicolumn{3}{c}{MIRRec} \\
\cline{2-16}
 & Top-1 & Top-3 & Top-5 & Top-1 & Top-3 & Top-5 & Top-1 & Top-3 & Top-5 & Top-1 & Top-3 & Top-5 & Top-1 & Top-3 & Top-5 \\
    \hline
    Bitcoin & 31.1 & 42.8 & 45.4 & 24.5 & 35.8 & 39.3 & \cellcolor{gray!15}32.0  & \cellcolor{gray!15}43.4 & \cellcolor{gray!15}46.4 & \cellcolor{gray!35}34.0  & \cellcolor{gray!35}46.3 & \cellcolor{gray!35}48.8 & \cellcolor{gray!60}\textbf{34.8} & \cellcolor{gray!60}\textbf{47.3} & \cellcolor{gray!60}\textbf{49.9} \\
    
    Electron & \cellcolor{gray!15}36.4 & 53.4 & \cellcolor{gray!15}56.9 & 16.3 & 31.0  & 37.1 & \cellcolor{gray!60}\textbf{41.7} & 56.9 & \cellcolor{gray!15}60.7 & \cellcolor{gray!60}\textbf{41.7} & \cellcolor{gray!35}57.2 & \cellcolor{gray!35}60.8 & \cellcolor{gray!35}41.5 & \cellcolor{gray!60}\textbf{57.4} & \cellcolor{gray!60}\textbf{60.9} \\
    
    Opencv & 58.9 & \cellcolor{gray!15}70.3 & 71.9 & 47.7 & 60.0  & 62.0  & \cellcolor{gray!60}\textbf{69.4} & \cellcolor{gray!35}77.8 & \cellcolor{gray!35}78.9 & \cellcolor{gray!15}65.3 & 76.1 & \cellcolor{gray!15}77.5 & \cellcolor{gray!35}68.8 & \cellcolor{gray!60}\textbf{78.2} & \cellcolor{gray!60}\textbf{79.6} \\
    
    XBMC & 29.0  & 40.5 & 43.3 & 34.0  & 45.1 & 47.9 & \cellcolor{gray!15}42.8 & \cellcolor{gray!15}52.0  & \cellcolor{gray!15}54.2 & \cellcolor{gray!35}43.2 & \cellcolor{gray!35}53.4 & \cellcolor{gray!35}55.8 & \cellcolor{gray!60}\textbf{44.9} & \cellcolor{gray!60}\textbf{55.4} & \cellcolor{gray!60}\textbf{57.7} \\
    
    React & 24.3 & 38.0  & 42.3 & 22.2 & 36.3 & 40.0  & \cellcolor{gray!35}43.6 & \cellcolor{gray!60}\textbf{57}  & \cellcolor{gray!35}59.6 & \cellcolor{gray!15}41.7 & 55.5 & \cellcolor{gray!15}58.6 & \cellcolor{gray!60}\textbf{43.8} & \cellcolor{gray!35}56.9 & \cellcolor{gray!60}\textbf{60} \\
    
    Angular & 29.6 & 43.5 & 47.2 & 32.6 & 46.1 & 49.3 & \cellcolor{gray!15}40.2 & \cellcolor{gray!15}51.6 & \cellcolor{gray!15}53.9 & \cellcolor{gray!35}46.8 & \cellcolor{gray!35}58.0  & \cellcolor{gray!35}60.6 & \cellcolor{gray!60}\textbf{48.5} & \cellcolor{gray!60}\textbf{59.6} & \cellcolor{gray!60}\textbf{62.0} \\
    
    Django & 46.9 & 61.8 & 63.8 & 36.0  & 50.8 & 53.6 & \cellcolor{gray!15}57  & \cellcolor{gray!15}66.2 & \cellcolor{gray!15}67.9 & \cellcolor{gray!35}59.1 & \cellcolor{gray!35}69.5 & \cellcolor{gray!35}70.9 & \cellcolor{gray!60}\textbf{59.9} & \cellcolor{gray!60}\textbf{70.0} & \cellcolor{gray!60}\textbf{71.4} \\

    Symfony & \cellcolor{gray!15}46.8 & \cellcolor{gray!15}61.7 & \cellcolor{gray!15}63.4 & 36.8 & 50.6 & 53.4 & 45.4 & 60.0  & 62.0  & \cellcolor{gray!35}48.2 & \cellcolor{gray!35}62.8 & \cellcolor{gray!35}64.8 & \cellcolor{gray!60}\textbf{49.5} & \cellcolor{gray!60}\textbf{64.2} & \cellcolor{gray!60}\textbf{65.9} \\
    
    Rails & \cellcolor{gray!60}\textbf{24.7} & \cellcolor{gray!60}\textbf{33.4} & \cellcolor{gray!60}\textbf{36.5} & 14.7 & 23.2 & 26.2 & 20.9 & 29.9 & 32.8 & \cellcolor{gray!15}21.9 & \cellcolor{gray!15}31.8 & \cellcolor{gray!15}34.9 & \cellcolor{gray!35}22.3 & \cellcolor{gray!35}33.2 & \cellcolor{gray!35}36 \\
    
    Scala & 43.1 & 56.6 & 59.4 & 28.8 & 43.6 & 46.9 & \cellcolor{gray!15}45.5 & \cellcolor{gray!15}59.7 & \cellcolor{gray!15}61.9 & \cellcolor{gray!35}46.9 & \cellcolor{gray!35}61.7 & \cellcolor{gray!35}64.2 & \cellcolor{gray!60}\textbf{48.3} & \cellcolor{gray!60}\textbf{62.7} & \cellcolor{gray!60}\textbf{65.2} \\
    \hline
    
    Average & 37.1 & 50.2 & 53  & 29.4 & 42.3 & 45.6 & \cellcolor{gray!15}43.9 & \cellcolor{gray!15}55.5 & \cellcolor{gray!15}57.8 & \cellcolor{gray!35}44.9 & \cellcolor{gray!35}57.2 & \cellcolor{gray!35}59.7 & \cellcolor{gray!60}\textbf{46.2} & \cellcolor{gray!60}\textbf{58.5} & \cellcolor{gray!60}\textbf{60.9} \\
    \hline
    
    Improve & 24.53\% & 16.53\% & 14.91\% & 57.14\% & 38.30\% & 33.55\% & 5.24\% & 5.41\% & 5.36\% & 2.90\% & 2.27\% & 2.01\% &  - & -  & - \\
    \hline
    \multicolumn{16}{c}{\textbf{Note}: `Average', `Improve', font and colors have the same meanings as in Table \ref{table4}.} \\
  \hline
  \end{tabularx}%
  \label{tabel5}%
\end{table*}%

Table \ref{table4} and \ref{tabel5} display the overall performance of MIRRec and four baseline approaches on the dataset. In terms of ACC, MIRRec outperforms most of the baselines on the majority of projects. Across all projects, MIRRec consistently achieves higher Top-1, Top-3, and Top-5 recommendation average ACC metrics. Specifically, MIRRec outperforms RevFinder by 24.53\%, 10.63\%, and 6.45\%, respectively; cHRev by  57.14\%, 25.04\%, and 14.56\%; CN by  5.24\%, 5.42\%, and 4.60\%; and HGRec by 2.90\%, 1.79\%, and 1.08\%. MRR is an indicator that reflects the order of recommendation results. The larger the MRR, the higher the average ranking of recommended real code reviewers. Our results for MRR mirror those of ACC, with MIRRec consistently leading the performance among all the recommenders in this study. As shown in Table \ref{tabel5}, MIRRec achieves, on average,  24.53\%, 16.53\%, and 14.91\% higher MRR values for  Top-$1$, Top-$3$, and Top-$5$ compared to RevFinder; 57.14\%, 38.30\%, and 33.55\%  higher than cHRev; 5.24\%, 5.41\%, and 5.36\% higher than CN; and  2.90\%, 2.27\%, and 2.01\%higher than HGRec. \textbf{These results demonstrate that MIRRec can accurately recommend the first real reviewers at a lower rank than the baselines, highlighting its superior recommendation accuracy}.


The potential reasons behind this phenomenon are summarized as follows:
$(1)$ MIRRec, CN, and HGRec are graph-based algorithms that excel at conducting feature learning for developers and PRs by exploring explicit and implicit multi-relationships, thereby significantly enhancing recommendation accuracy.
$(2)$ Compared to graph-based methods, hypergraph-based methods, such as HGRec and MIRRec, have distinct advantages in capturing higher-order relationships between nodes, leading to improved accuracy over graph-based methods like CN. Specifically, MIRRec is designed to capture potential correlation features between PRs and reviewers influenced by multiple relationships. This makes it superior in terms of accuracy compared to the HGRec method, which only considers PR-Reviewer and PR-Contributor relationships in interaction. 
$(3)$  MIRRec learns the representation of developers and PRs through interaction and similarity relationships. From the perspective of developers, MIRRec learns their representation from the tasks (PRs) they interact with. This is similar to how RevFinder learns developers’ expertise from the similarity of file path sets in tasks, or how cHRev learns developers’ expertise from their code files. However, unlike cHRev and RevFinder, MIRRec can extract more useful information and enhance feature learning by aggregating high-order neighbors.
\begin{mdframed}[backgroundcolor=gray!20]
\textbf{Answer}: MIRRec outperforms the baselines in ACC and MRR, demonstrating its accuracy and superiority in leveraging multi-relationships and higher-order connections between developer and PR for recommendation.
\end{mdframed}

\subsection{Results for RQ3: Impact of Relations}

\begin{table*}[hb]
  \centering
  \vspace{-0.6cm}
  \caption{ACC for Variants of MIRRec Based on Different Relations (\%)}
  \vspace{-0.2cm}
    \renewcommand{\arraystretch}{1.1}
    \begin{tabularx}{\linewidth}{p{0.7cm}*{9}{>{\centering\arraybackslash}X}p{0.5cm}p{0.5cm}p{0.5cm}*{9}{>{\centering\arraybackslash}X}}
    \hline
    \multirow{2}{*}{Project} & \multicolumn{3}{c}{MIRRec$_{ct\_ic\_rc}$} & \multicolumn{3}{c}{MIRRec$_{re\_ic\_rc}$} & \multicolumn{3}{c}{MIRRec$_{re\_ct\_rc}$} & \multicolumn{3}{c}{MIRRec$_{re\_ct\_ic}$} & \multicolumn{3}{c}{MIRRec$_{re\_ct}$}& \multicolumn{3}{c}{MIRRec$_{re\_ic}$}& \multicolumn{3}{c}{MIRRec} \\
\cline{2-22}
 & \mbox{Top-1} & \mbox{Top-3} & \mbox{Top-5} & \mbox{Top-1} & \mbox{Top-3} & \mbox{Top-5} & \mbox{Top-1} & \mbox{Top-3} & \mbox{Top-5} & \mbox{Top-1} & \mbox{Top-3} & \mbox{Top-5} & \mbox{Top-1} & \mbox{Top-3} & \mbox{Top-5} & \mbox{Top-1} & \mbox{Top-3} & \mbox{Top-5} & \mbox{Top-1} & \mbox{Top-3} & \mbox{Top-5}  \\
    \hline
    Bitcoin & 34.5 & 59.6 & 70.8 & 34.1 & 61.3 & 73.7 & \cellcolor{gray!35}35.1 & \cellcolor{gray!35}63.6 & \cellcolor{gray!15}74.4 & \cellcolor{gray!15}34.9 & \cellcolor{gray!15}63.5 & \cellcolor{gray!60}\textbf{75.1} & \cellcolor{gray!60}\textbf{35.5} & \cellcolor{gray!60}\textbf{63.9} & \cellcolor{gray!35}74.8 & 34.2 & 61.8 & 74.0 & 34.8 & 63.2 & \cellcolor{gray!35}74.8  \\
    
    Electron  & 33.8 & 71.7 & 90.3 & 41.4 & 77.0  & 92.7 & \cellcolor{gray!35}41.7 & \cellcolor{gray!15}78.0  & \cellcolor{gray!15}92.9 & \cellcolor{gray!35}41.7 & \cellcolor{gray!35}78.3 & \cellcolor{gray!60}\textbf{93.1} & \cellcolor{gray!60}\textbf{41.8} & \cellcolor{gray!60}\textbf{78.4} & \cellcolor{gray!35}93.0  & \cellcolor{gray!35}41.7 & 77.2 & \cellcolor{gray!35}93.0 & \cellcolor{gray!15}41.5 & \cellcolor{gray!15}78.0 & \cellcolor{gray!60}\textbf{93.1} \\
    
    Opencv  & 61.7 & 88.8 & 94.4 & 67.9 & \cellcolor{gray!35}89.7 & \cellcolor{gray!35}95.1 & 67.5 & \cellcolor{gray!15}89.6 & \cellcolor{gray!15}95  & \cellcolor{gray!60}\textbf{68.8} & \cellcolor{gray!35}89.7 & \cellcolor{gray!60}\textbf{95.5} & \cellcolor{gray!15}68.1 & \cellcolor{gray!35}89.7 & \cellcolor{gray!35}95.1 & \cellcolor{gray!35}68.4 & \cellcolor{gray!35}89.7 & \cellcolor{gray!35}95.1 & \cellcolor{gray!60}\textbf{68.8} & \cellcolor{gray!60}\textbf{89.8} & \cellcolor{gray!60}\textbf{95.5}\\
    
    XBMC  & 40.4 & 63.6 & 76.5 & 43.4 & 66.9 & 77.4 & 44.7 & 68.3 & \cellcolor{gray!35}79.0  & \cellcolor{gray!35}45.0  & \cellcolor{gray!15}68.5 & \cellcolor{gray!15}78.7 & \cellcolor{gray!60}\textbf{45.3} & \cellcolor{gray!60}\textbf{68.8} & \cellcolor{gray!60}\textbf{79.1} & 43.6 & 66.8 & 77.5& \cellcolor{gray!15}44.9 & \cellcolor{gray!35}68.7 & \cellcolor{gray!15}78.7 \\
    
    React  & 42.0  & 71.3 & 84.8 & 42.6 & 73.5 & 86.7 & \cellcolor{gray!15}43.6 & \cellcolor{gray!35}74.2 & \cellcolor{gray!35}87.0  & \cellcolor{gray!60}\textbf{43.8} & \cellcolor{gray!35}74.2 & \cellcolor{gray!60}\textbf{87.1} & \cellcolor{gray!35}43.7 & \cellcolor{gray!60}\textbf{74.7} & \cellcolor{gray!15}86.9 & 42.7 & 73.6 & 86.5 & \cellcolor{gray!60}\textbf{43.8} & \cellcolor{gray!15}73.8 & \cellcolor{gray!60}\textbf{87.1}\\
    
    Angular  & 44.8 & 70.6 & 82.3 & 47.2 & 72.8 & 83.6 & 48.1 & 73.3 & \cellcolor{gray!15}84.1 & \cellcolor{gray!60}\textbf{48.7} & \cellcolor{gray!60}\textbf{73.9} & \cellcolor{gray!60}\textbf{84.3} & \cellcolor{gray!35}48.6 & \cellcolor{gray!15}73.5 & \cellcolor{gray!35}84.2 & 47.2 & 72.9 & 83.6 & \cellcolor{gray!15}48.5 & \cellcolor{gray!35}73.8 & \cellcolor{gray!15}84.1\\
    
    Django  & 58.9 & 80.2 & 86.4 & 59.8 & \cellcolor{gray!15}81.8 & 87.9 & \cellcolor{gray!35}60.0  & \cellcolor{gray!15}81.8 & \cellcolor{gray!35}88.1 & \cellcolor{gray!60}\textbf{60.2} & \cellcolor{gray!35}81.9 & \cellcolor{gray!15}88.0  & \cellcolor{gray!60}\textbf{60.2} & \cellcolor{gray!35}81.9 & \cellcolor{gray!60}\textbf{88.2} & 59.6 & \cellcolor{gray!60}\textbf{82.0} & 87.9 & \cellcolor{gray!15}59.9 & \cellcolor{gray!35}81.9 & \cellcolor{gray!35}88.1\\
    
    Symfony & 45.6 & 80.1 & 88.2 & \cellcolor{gray!15}49.6 & 81.7 & 89.3 & 48.4 & 81.5 & \cellcolor{gray!15}89.5 & \cellcolor{gray!60}\textbf{50.5} & \cellcolor{gray!60}\textbf{82.4} & \cellcolor{gray!35}89.6 & 48.9 & 81.8 & \cellcolor{gray!60}\textbf{89.7} & \cellcolor{gray!35}49.8 & \cellcolor{gray!15}81.9 & \cellcolor{gray!15}89.5 & 49.5 & \cellcolor{gray!35}82.2 & \cellcolor{gray!15}89.5 \\
    
    Rails  & 16.6 & 42.2 & 56.3 & \cellcolor{gray!35}22.7 & \cellcolor{gray!35}47.1 & 59.0  & 21.1 & 46.3 & \cellcolor{gray!15}59.4 & \cellcolor{gray!15}22.6 & \cellcolor{gray!35}47.1 & \cellcolor{gray!60}\textbf{60.0} & 21.2 & 46.2 & 59.0  & \cellcolor{gray!60}\textbf{22.8} & \cellcolor{gray!15}46.9 & 59.1 & 22.3 & \cellcolor{gray!60}\textbf{47.4} & \cellcolor{gray!35}59.6\\
    
    Scala  & 43.5 & 78.4 & 89.8 & 47.2 & \cellcolor{gray!15}80.0  & 90.4 & 47.4 & \cellcolor{gray!35}80.3 & \cellcolor{gray!15}90.7 & \cellcolor{gray!35}48.1 & \cellcolor{gray!60}\textbf{80.4} & \cellcolor{gray!60}\textbf{91.3} & \cellcolor{gray!15}47.5 & \cellcolor{gray!60}\textbf{80.4} & 90.8 & 47.4 & \cellcolor{gray!15}80.0  & 90.3 & \cellcolor{gray!60}\textbf{48.3} & \cellcolor{gray!35}\cellcolor{gray!35}80.3 & \cellcolor{gray!35}91.0\\
    \hline
    
    Average  & 42.2 & 70.7 & 82.0  & 45.6 & 73.2 & 83.6 & 45.8 & \cellcolor{gray!15}73.7 & 84.0  & \cellcolor{gray!60}\textbf{46.4} & \cellcolor{gray!60}\textbf{74.0} & \cellcolor{gray!60}\textbf{84.3} & \cellcolor{gray!15}46.1 & \cellcolor{gray!35}73.9 & \cellcolor{gray!15}84.1 & 45.7 & 73.3 & 83.7 & \cellcolor{gray!35}46.2 & \cellcolor{gray!35}73.9 & \cellcolor{gray!35}84.2\\
    \hline
    
    Improve   & 9.48\% & 4.53\% & 2.68\% & 1.32\% & 0.96\% & 0.72\% & 0.87\% & 0.27\% & 0.24\% &  \mbox{-0.43}\% & \mbox{-0.14}\% & \mbox{-0.12}\% & 0.22\% &- & 0.12\% & 1.09\% & 0.82\% & 0.60\%  & -   & -   & -\\
    \hline
    \multicolumn{22}{c}{\textbf{Note}: `Average', `Improve', font and colors have the same meanings as in Table \ref{table4}.} \\
    \hline 
  \end{tabularx}
 \label{tabel6}%
 \end{table*}
 \vspace{-0.1cm}

\begin{table*}[bp]
  \centering
  \vspace{-0.3cm}
  \caption{MRR for Variants of MIRRec Based on Different Relations (\%) }
  \vspace{-0.2cm}
    \renewcommand{\arraystretch}{1.1}
    \begin{tabularx}{\linewidth}{p{0.7cm}*{9}{>{\centering\arraybackslash}X}p{0.5cm}p{0.5cm}p{0.5cm}*{9}{>{\centering\arraybackslash}X}}
    \hline
    \multirow{2}{*}{Project} & \multicolumn{3}{c}{MIRRec$_{ct\_ic\_rc}$} & \multicolumn{3}{c}{MIRRec$_{re\_ic\_rc}$} & \multicolumn{3}{c}{MIRRec$_{re\_ct\_rc}$} & \multicolumn{3}{c}{MIRRec$_{re\_ct\_ic}$} & \multicolumn{3}{c}{MIRRec$_{re\_ct}$}& \multicolumn{3}{c}{MIRRec$_{re\_ic}$}& \multicolumn{3}{c}{MIRRec} \\
\cline{2-22}
 & \mbox{Top-1} & \mbox{Top-3} & \mbox{Top-5} & \mbox{Top-1} & \mbox{Top-3} & \mbox{Top-5} & \mbox{Top-1} & \mbox{Top-3} & \mbox{Top-5} & \mbox{Top-1} & \mbox{Top-3} & \mbox{Top-5} & \mbox{Top-1} & \mbox{Top-3} & \mbox{Top-5} & \mbox{Top-1} & \mbox{Top-3} & \mbox{Top-5} & \mbox{Top-1} & \mbox{Top-3} & \mbox{Top-5}  \\
    \hline
    Bitcoin & 34.5 & 45.5 & 48.1 & 34.1 & 46.1 & 49.0  & \cellcolor{gray!35}35.1 & \cellcolor{gray!35}47.6 & \cellcolor{gray!35}50.1 & \cellcolor{gray!15}34.9 & \cellcolor{gray!15}47.5 & \cellcolor{gray!35}50.1 & \cellcolor{gray!60}\textbf{35.5} & \multicolumn{1}{r}{\cellcolor{gray!60}\textbf{48.0}} & \multicolumn{1}{r}{\cellcolor{gray!60}\textbf{50.5}} & 34.2 & \multicolumn{1}{r}{46.3} & \multicolumn{1}{r}{49.1} & 34.8 & 47.3 & \cellcolor{gray!15}49.9 \\
    
    Electron & 33.8 & 50.1 & 54.4 & 41.4 & 57.0  & 60.7 & \cellcolor{gray!35}41.7 & \cellcolor{gray!15}57.5 & \cellcolor{gray!15}60.9 & \cellcolor{gray!35}41.7 & \cellcolor{gray!35}57.7 & \cellcolor{gray!35}61.1 & \cellcolor{gray!60}\textbf{41.8} & \cellcolor{gray!60}\textbf{57.8} & \cellcolor{gray!60}\textbf{61.2} & \cellcolor{gray!35}41.7 & 57.3 & \cellcolor{gray!15}60.9 & \cellcolor{gray!15}41.5 & 57.4 & \cellcolor{gray!15}60.9 \\
    
    Opencv & 61.7 & 74.0  & 75.3 & 67.9 & 77.8 & 79.0  & 67.5 & 77.6 & 78.9 & \cellcolor{gray!60}\textbf{68.8} & \cellcolor{gray!60}\textbf{78.3} & \cellcolor{gray!60}\textbf{79.6} & \cellcolor{gray!15}68.1 & 77.9 & \cellcolor{gray!15}79.1 & \cellcolor{gray!35}68.4 & \cellcolor{gray!15}78.0  & \cellcolor{gray!35}79.3 & \cellcolor{gray!60}\textbf{68.8} & \cellcolor{gray!35}78.2 & \cellcolor{gray!60}\textbf{79.6} \\
    
    XBMC & 40.4 & 50.5 & 53.5 & 43.4 & 53.6 & 56.0  & 44.7 & \cellcolor{gray!15}55.1 & \cellcolor{gray!15}57.6 & \cellcolor{gray!35}45.0  & \cellcolor{gray!35}55.4 & \cellcolor{gray!35}57.7 & \cellcolor{gray!60}\textbf{45.3} & \cellcolor{gray!60}\textbf{55.6} & \cellcolor{gray!60}\textbf{58.0} & 43.6 & 53.7 & 56.2 & \cellcolor{gray!15}44.9 & \cellcolor{gray!35}55.4 & \cellcolor{gray!35}57.7 \\
    
    React & 42.0  & 54.8 & 58.0  & 42.6 & 56.2 & \cellcolor{gray!15}59.2 & \cellcolor{gray!15}43.6 & \cellcolor{gray!15}57.0  & 60.0  & \cellcolor{gray!60}\textbf{43.8} & \cellcolor{gray!35}57.1 & \cellcolor{gray!60}\textbf{60.1} & \cellcolor{gray!35}43.7 & \cellcolor{gray!60}\textbf{57.2} & \cellcolor{gray!35}60.0  & 42.7 & 56.2 & \cellcolor{gray!15}59.2 & \cellcolor{gray!60}\textbf{43.8} & 56.9 & \cellcolor{gray!35}60.0 \\
    
    Angular & 44.8 & 56.2 & 58.9 & 47.2 & 58.6 & 61.0  & 48.1 & 59.2 & 61.6 & \cellcolor{gray!60}\textbf{48.7} & \cellcolor{gray!60}\textbf{59.8} & \cellcolor{gray!60}\textbf{62.2} & \cellcolor{gray!35}48.6 & \cellcolor{gray!15}59.5 & \cellcolor{gray!15}61.9 & 47.2 & 58.7 & 61.1 \cellcolor{gray!15}& 48.5 & \cellcolor{gray!35}59.6 & \cellcolor{gray!35}62.0 \\
    
    Django & 58.9 & 68.4 & 69.8 & 59.8 & \cellcolor{gray!15}69.8 & \cellcolor{gray!15}71.2 & \cellcolor{gray!35}60.0  & \cellcolor{gray!35}70.0  & \cellcolor{gray!35}71.4 & \cellcolor{gray!60}\textbf{60.2} & \cellcolor{gray!60}\textbf{70.1} & \cellcolor{gray!60}\textbf{71.5} & \cellcolor{gray!60}\textbf{60.2} & \cellcolor{gray!60}\textbf{70.1} & \cellcolor{gray!60}\textbf{71.5} & 59.6 & \cellcolor{gray!15}69.8 & \cellcolor{gray!15}71.2 & \cellcolor{gray!15}59.9 & \cellcolor{gray!35}70.0  & \cellcolor{gray!35}71.4 \\
    
    Symfony & 45.6 & 61.2 & 63.1 & \cellcolor{gray!15}49.6 & 64.1 & 65.8 & 48.4 & 63.3 & 65.2 & \cellcolor{gray!60}\textbf{50.5} & \cellcolor{gray!60}\textbf{64.8} & \cellcolor{gray!60}\textbf{66.4} & 48.9 & 63.7 & 65.5 & \cellcolor{gray!35}49.8 & \cellcolor{gray!35}64.3 & \cellcolor{gray!35}66.0  & 49.5 & \cellcolor{gray!15}64.2 & \cellcolor{gray!15}65.9 \\
    
    Rails & 16.6 & 27.6 & 30.8 & \cellcolor{gray!35}22.7 & \cellcolor{gray!60}\textbf{33.3} & 36.0  & 21.1 & 32.2 & 35.2 & \cellcolor{gray!15}22.6 & \cellcolor{gray!60}\textbf{33.3} & \cellcolor{gray!60}\textbf{36.3} & 21.2 & \cellcolor{gray!15}32.3 & 35.2 & \cellcolor{gray!60}\textbf{22.8} & \cellcolor{gray!60}\textbf{33.3} & \cellcolor{gray!35}36.1 & 22.3 & \cellcolor{gray!35}33.2 & \cellcolor{gray!15}36.0 \\
    
    Scala & 43.5 & 58.9 & 61.6 & 47.2 & 62.0  & 64.4 & 47.4 & \cellcolor{gray!35}62.4 & \cellcolor{gray!35}64.8 & \cellcolor{gray!35}48.1 & \cellcolor{gray!60}\textbf{62.7} & \cellcolor{gray!60}\textbf{65.2} & \cellcolor{gray!15}47.5 & \cellcolor{gray!35}62.4 & \cellcolor{gray!35}64.8 & 47.4 & \cellcolor{gray!15}62.1 & \cellcolor{gray!15}64.5 & \cellcolor{gray!60}\textbf{48.3} & \cellcolor{gray!60}\textbf{62.7} & \cellcolor{gray!60}\textbf{65.2} \\
    
    Avarage & 42.2 & 54.7 & 57.4 & 45.6 & 57.9 & 60.2 & 45.8 & \cellcolor{gray!15}58.2 & 60.6 & \cellcolor{gray!60}\textbf{46.4} & \cellcolor{gray!60}\textbf{58.7} & \cellcolor{gray!60}\textbf{61.0} & \cellcolor{gray!15}46.1 & \cellcolor{gray!35}58.5 & \cellcolor{gray!15}60.8 & 45.7 & 58.0  & 60.4 & \cellcolor{gray!35}46.2 & \cellcolor{gray!35}58.5 & \cellcolor{gray!35}60.9 \\
    \hline
    Improve & 9.48\% & 6.95\% & 6.10\% & 1.32\% & 1.04\% & 1.16\% & 0.87\% & 0.52\% & 0.50\% & \mbox{-0.43\%} & \mbox{-0.34\%} & \mbox{-0.16\%} & 0.22\% & - & 0.16\% & 1.09\% & 0.86\% & 0.83\% & - & - &  -\\
    \hline
    \multicolumn{22}{c}{\textbf{Note}: `Average', `improve', font and colors have the same meanings as in Table \ref{table4}.} \\
    \hline
    \end{tabularx}
    \label{tabel7}%
    \end{table*}
    
To further investigate the impact of different hyperedges and the effectiveness of MIRRec based on the hypergraph models constructed with different hyperedges, we constructed several models by pruning the relationships, including $\text{MIRRec}_{ct\_ic\_rc}$, $\text{MIRRec}_{re\_ic\_rc}$, $\text{MIRRec}_{re\_ct\_rc}$, $\text{MIRRec}_{re\_ct\_ic}$, $\text{MIRRec}_{re\_ct}$, and $\text{MIRRec}_{re\_ic}$, using the weight combination setting mentioned above. The subscript represents the partial interaction relationships used to construct each model: $re$ (PR-Reviewers), $ct$ (PR-Committers), $ic$ (PR-Issue Commenters), and $rc$ (PR-Review Commenters). 
It is worth noting that we did not perform an ablation experiment on the PR-creator relationship. The reason is that the creator is the closest role to the PR compared to the reviewer and commenter, reflecting developers' involvement and expertise in the relevant domain of the PR under review. This observation aligns with the phenomenon we discovered, that many creators review their own created PRs, supporting the significance of considering this relationship as fundamental.
Tables \ref{tabel6} and \ref{tabel7} display the performance of these models in terms of ACC and MRR, respectively.

As shown in Table \ref{tabel6}, MIRRec achieves an average Top-1 ACC that is approximately 9.48\% higher than $\text{MIRRec}_{ct\_ic\_rc}$. This reveals the significant impact of the $re$ relationship on recommendations, aligning with the regularity of recommendation systems based on historical interactions. MIRRec also slightly outperforms $\text{MIRRec}_{re\_ic\_rc}$ with an improvement of 1.32\%, 0.96\%, 0.72\% for Top-$1$, Top-$3$, and Top-$5$, indicating a positive influence from the $ct$ relationship, likely due to its ability to capture familiarity with PRs and potential developer collaboration. While the ACC of MIRRec is slightly lower than that of $\text{MIRRec}_{re\_ct}$ on certain projects, the average ACC remains higher for MIRRec than for $\text{MIRRec}_{re\_ct}$. This suggests that the two comment relationships, $ic$ and $rc$, may have some potential positive impact but also carry the risk of complex information disruptions. Additional experiments with $\text{MIRRec}_{re\_ct\_rc}$, $\text{MIRRec}_{re\_ct\_ic}$, and $\text{MIRRec}_{re\_ic}$ also confirm the potential risk of information interference.

Meanwhile, the performance of the MRR of these models shows a high similar to the performance of ACC as Table \ref{tabel7} shows. From these results we can infer that $ic$ usually plays a positive role, while $rc$ has a negative impact on some projects. We speculate that this may be because $rc$ is generated based on $re$, and thus it may introduce some invalid or redundant information on top of the already existing review interactions. The relation $ic$, on the other hand, can somewhat capture the potential review and collaboration possibilities of other developers as opposed to the review comment relationship. 
Therefore, we recommend using MIRRec without the $rc$ relationship as the optimal recommender.

We further analyze the statistics to find the potential reasons for these phenomena. In the interaction data related to $ct$, $ic$, $re$, and $rc$ of the datasets, we observe that about 76\% to 87\% of developers in the projects played only one role out of $ct$, $ic$, and $re$. Moreover, 7\% to 15\% took on two of the four roles, 3\% to 7\% managed three of the four roles, and 2\% to 7\% juggled all four roles simultaneously. It should be noted that many developers who served as $rc$ also took on other roles within the project. In the Bitcoin, React, and Rails projects, developers exclusively held the role of $rc$, and this accounted for only about 1\% of the total. This highlights the importance of learning higher-order relationships between developers and PRs from the interaction histories related to $ct$, $ic$, and $re$ data. However, it is important to note that the data from $rc$, in addition to $ct$, $ic$, and $re$, can be considered redundant since these three categories already encompass most of the $rc$ data in the projects. This redundancy may lead to overfitting.

\begin{mdframed}[backgroundcolor=gray!20]
\textbf{Answer}: The $re$ relationship significantly influences recommendations, with $ct$ and $ic$ contributing positively, while $rc$ may introduce redundancy. This underscores the importance of considering higher-order relationships but also raises caution about potential overfitting due to overlapping data, highlighting the need for meticulous data analysis in recommendation systems.
\end{mdframed}
\section{Threats to validaity}\label{section5}

\textbf{External Validity}. 
The evaluation experiments are conducted on a dataset we constructed rather than on previously publicly available datasets, such as the dataset introduced by Rong et al. \cite{b22}. The reason is that the available datasets released may not be sourced directly and lack specific relationships. Consequently, obtaining the necessary multiplex relational information proves unfeasible.
While the projects in our dataset may not represent all OSS projects,
they were selected based on predetermined criteria to ensure their relevance and diversity.  Furthermore, some of the projects utilized have been referenced in previous studies and the data is up-to-date. As part of our future work, we plan to expand the evaluation to a wider range of projects, both open-source and industrial.

\textbf{Internal Validity}. The internal validity of this work is primarily related to data preprocessing, experiment setup, and hyperparameter tuning. During data preprocessing, issues such as inconsistent developer names and invalid identities were addressed, which could have disrupted the accurate representation of interaction relationships. To mitigate these threats, we applied various preprocessing techniques to construct the dataset. In terms of experimental setup, we partitioned the dataset into sliding training and testing subsets for multiple rounds of experimentation, using average performance as the evaluation metric. This approach helps mitigate threats arising from varying data sample sizes due to different project development velocities over time. Regarding hyperparameter tuning, while it is challenging to determine optimal values exhaustively, we conducted a series of `trial and error' experiments to identify suitable hyperparameters. This approach effectively reduces the potential threats associated with suboptimal hyperparameters to a certain extent.

\textbf{Construct Validity}. MIRRec is a recommendation approach that focuses on relationship-aware reviewer recommendations using a hypergraph of multiplex relationships. To construct this hypergraph, we implemented a series of data preprocessing steps to retain crucial information and minimize the impact of redundant data, enhancing the interpretability of our research findings. We evaluated the effectiveness of our method by comparing it with four state-of-the-art baselines using metrics such as ACC and MRR. 
In terms of workload assessment for the recommended reviewers of MIRRec, we have not delved into detailed discussions in this paper. However, we provide specific results on the Recommendation Distribution \cite{b22} of MIRRec and baseline methods in the replication package~\cite{b27}. Our study results indicate that, while maintaining excellent accuracy, MIRRec, in contrast to other baseline methods, is more adept at considering the balanced distribution of workload, thus providing more diverse recommendation outcomes.
The results from our ablation experiments further underscore the efficacy of constructing a hypergraph based on multiplex relationships. However, it is worth noting that our approach relies on the actual reviewers assigned to tasks as ground truth, without considering factors such as their expertise, reputation, and workload. This limitation could potentially lead to recommendations that may not always be the most suitable. 

\textbf{Conclusion Validity}. To ensure the validity of our conclusions, we followed a meticulous and systematic procedure for experimentation and analysis. Our proposed method was tested on ten OSS projects with rich interaction histories, encompassing over $4.8$K PRs and $47.3$K interactions. We have provided a clear elaboration of all the data used, which is publicly available. This transparency enhances the reliability of our conclusions, as they can be traced back to the original data, allowing for replication by other researchers.

\section{Related work}\label{section6}
Industry and academia have made great efforts in recommending code reviewers\cite{b26} recently, and put forward many approaches to select the appropriate reviewers.  

As the necessity for expertise and prior knowledge in code review, many studies have noted the importance of recommending suitable reviewers, which can lead to faster and higher quality software updates. For instance, Hannebauer et al.\cite{b29} found that the code reviewers’ recommendation algorithms based on review expertise outperform those based solely on modification expertise. Asthana et al. \cite{b30} identified potential reviewers by leveraging their past experience with the files and directories involved in a code review. Fejzer et al. \cite{b31} suggested reviewers based on the similarity between developers’ expertise profiles and changes to be reviewed. Rahman et al.\cite{b32} heuristically captured relevant cross-project work history and specialized technologies used in a PR for reviewer recommendation. However, these methods made solely based on developers’ individual qualities may result in a focus on recommending core reviewers. In response to this, Rebai et al. \cite{b33} formulated reviewer recommendation as a multi-objective search problem by constructing an expertise model, availability model, and a collaboration model to balance the conflicting objectives of expertise, availability, and history of collaborations. Some researchers\cite{b5}\cite{b34} employed multi-objective search-based method to find the optimal reviewers by balancing both expertise and workload. 

Most of the reviewer recommenders that we know depend on historical reviews information, and recommend developers based on matrix factorization or collaborative filtering \cite{b35}\cite{b36}\cite{b37}. For example, Chueshev et al. \cite{b37} introduced a recommender-based approach for OSS projects, which uses collaborative filtering to recommend regular reviewers and expand their numbers from suitable developers. Some researchers have acknowledged the value of additional information in update changes to enhance reviewer recommendation. Xia et al.\cite{b38} proposed a hybrid and incremental approach to recommend code reviewers, combining text mining and file location-based similarity measures between new and previous files. Ye et al. \cite{b39} proposed a multi-instance-based deep neural network model that utilizes LSTM and CNN to recommend reviewers for PRs based on information from the PR title, commit message, and code changes. Li et al.\cite{b46} proposed an approach to recommend code reviewers for architecture violation issues based on the file path similarity of code commits and the semantic similarity of review comments. Jiang et al.\cite{b40} built approaches based on various attributes, including activeness, text similarity, file similarity and social relations to recommend commenters for PRs. Jiang et al.\cite{b41} also developed a classifier based on SVM that considers features such as file paths of modified codes, relationships between contributors and core members, and the activeness of core members to recommend core reviewers for PRs. While these methods that utilize historical and change information have been validated and successful, they may overlook capable reviewers who have not interacted with these tasks before.

Another line of research focuses more on leveraging graph techniques within the developers' network\cite{b42}\cite{b43}. Liao et al. \cite{b44} proposed a core-reviewer recommendation approach that combines PR topic model with collaborators in the social network. Ying et al. \cite{b19} constructed a graph connecting incoming PRs with potential reviewers, using text similarity of PRs and social relations among developers to identify suitable reviewers, effectively considering developer expertise and authority. Sülün et al. \cite{b20} introduced the ‘know-about’ metric to measure familiarity between developers and artifacts and utilized traceability graphs of software artifacts to recommend reviewers for specific changes. Zhang et al.\cite{b45}  employed a graph convolutional neural network for reviewer recommendation that leverages a socio-technical graph built from the rich set of entities (developers, repositories, files, PRs, work items, etc. Yu et al.\cite{b18} extended three typical approaches based on machine learning, information retrieval, and file location techniques to recommend reviewers to PRs. They also proposed a hybrid approach that combines the comment network with traditional approaches, achieving significant improvement over the traditional methods. Rong et al. \cite{b22} adopted the hypergraph to model high-order relationships between PRs and developers, then recommended reviewers based on the model. However, a common limitation of these is that they do not sufficiently consider the complex interactions between developers and tasks in the review process.

\textbf{Key difference}.
In contrast to prior work, our objective is to explore the multiplex relationships between developers and PRs, as well as their high-order connectivities in the review process, with the goal of providing more accurate reviewer recommendations. Unlike previous studies, which typically incorporate explicit historical review interactions (PR-Reviewer) and task similarity (PR-PR) into their models, and consider contributor interactions (eg. PR-Committer and PR-Issue Commenter) as a single type relationship beyond the review interaction. We take a different strategy treat these relationships separately. 
While HGRec made an improvement by using hypergraphs to model code reviews and the intricate relationships among participants,
our approach, MIRRec goes further by measuring different hyperedges using multiple factors, including commit time, lines of code modified, comment time, comment frequency, and the behavior recency of developers. Thus, emphasizing the depth and specificity of different contributions. 
In summary, MIRRec places a stronger emphasis on advancing the understanding and learning the different contributions of these complex relationships based on diverse behaviors under different roles. The evaluation results on OSS projects suggest that conducting an in-depth analysis of the implications of various relations and hypergraph technique contributes to the improved accuracy of recommendations.

\section{Conclusion and future work}\label{section7}
In this paper, we propose MIRRec, a novel relationship-aware hypergraph-based approach for enhancing reviewer recommendation in OSS projects. MIRRec formally defines the interaction relationships between developers and PRs and the similarity relationship between PRs. It constructs a multiplex-relationship hypergraph to model high-order relationships between developers and PRs. Comprehensive experiments conducted on ten OSS projects indicate that MIRRec outperforms state-of-the-art baselines. An ablation study was also carried out to investigate the influence of different relationships on reviewer recommendations, further validating the effectiveness of the proposed method.

Our results reveal that the majority of developers engage in separate types of interactions, including commits, reviews, and issue comments, within the project. Some developers exhibit at least two types of interaction behaviors, and the majority of developers who contribute review comments also participate in additional types of interactions. Only a tiny percentage of developers exclusively participate in review comment activities. Consequently, it is crucial to consider relationships related to commits and issue comments alongside the historical review interactions. It is worth noting that when the recommender aims to comprehensively learn the potential high-order connectivity based on the additional review comment behaviors, addressing the issue of data redundancy leading to overfitting is necessary.

In the future, we aim to enhance MIRRec by incorporating developers' individual abilities such as expertise, response quality, contributions, and impact into their representations. Additionally, we plan to expand our evaluation to a wider range of OSS projects and deploy MIRRec as a GitHub plugin to conduct practical experiments with developers to assess its effectiveness in reviewer recommendations.

\section*{Acknowledgements}
This work is supported by the National Natural Science Foundation of China (Nos. 62032016 and 62172311), the Key Research and Development Program of Hubei Province (No. 2021BAA031), and the Natural Science Foundation of Hubei Province (No. 2023AFB374).

\balance


\begin{thebibliography}{00}

\bibitem{b1} D. Badampudi, M. Unterkalmsteiner, and R. Britto, “Modern Code Reviews—Survey of Literature and Practice,” ACM Transactions on Software Engineering and Methodology, vol. 32, no. 4, pp. 1-61, 2023.

\bibitem{b2} X. Yang, R. G. Kula, N. Yoshida, and H. Iida, “Mining the modern code review repositories: a dataset of people, process and product,” in Proceedings of the 13th International Conference on Mining Software Repositories (MSR), pp. 460-463, 2016.

\bibitem{b3} S. Ruangwan, P. Thongtanunam, A. Ihara, and K. Matsumoto, “The impact of human factors on the participation decision of reviewers in modern code review,” Empirical Software Engineering, vol. 24, no. 2, pp. 973-1016, 2018.

\bibitem{b4} D. Kong, Q. Chen, L. Bao, C. Sun, X. Xia, and S. Li, “Recommending Code Reviewers for Proprietary Software Projects: A Large Scale Study,” in Proceedings of the 29th IEEE International Conference on Software Analysis, Evolution and Reengineering (SANER), pp. 630-640, 2022.

\bibitem{b5} M. Chouchen, A. Ouni, M. W. Mkaouer, R. G. Kula, and K. Inoue, “WhoReview: A multi-objective search-based approach for code reviewers recommendation in modern code review,” Applied Soft Computing, vol. 100, 106908, 2021.

\bibitem{b6} E. Doğan, E. Tüzün, K. A. Tecimer, and H. A. Güvenir, “Investigating the Validity of Ground Truth in Code Reviewer Recommendation Studies,” in Proceedings of the 13th ACM/IEEE International Symposium on Empirical Software Engineering and Measurement (ESEM), pp. 1-6, 2019.

\bibitem{b7} G. Zhao, D. A. da Costa, and Y. Zou, “Improving the pull requests review process using learning-to-rank algorithms,” Empirical Software Engineering, vol. 24, no. 4, pp. 2140-2170, 2019.

\bibitem{b8} P. C. Rigby, and C. Bird, “Convergent contemporary software peer review practices,” in Proceedings of the 9th Joint Meeting of the European Software Engineering Conference and the ACM SIGSOFT Symposium on the Foundations of Software Engineering (ESEC/FSE), pp. 202-212, 2013.

\bibitem{b9}J Tsay, L. Dabbish, and J. Herbsleb, “Let's Talk About It: Evaluating Contributions through Discussion in GitHub,” in Proceedings of the 22nd ACM SIGSOFT International Symposium on the Foundations of Software Engineering (FSE), pp. 144-154, 2014.

\bibitem{b10}A. Lee, J. C. Carver, and A. Bosu, “Understanding the Impressions, Motivations, and Barriers of One Time Code Contributors to FLOSS Projects: A Survey,” in Proceedings of the 39th IEEE/ACM International Conference on Software Engineering (ICSE), pp. 187-197, 2017.

\bibitem{b11}V. Balachandran, “Reducing human effort and improving quality in peer code reviews using automatic static analysis and reviewer recommendation,” in Proceedings of the 35th International Conference on Software Engineering (ICSE), pp. 931-940, 2013. 

\bibitem{b12}P. Thongtanunam, C. Tantithamthavorn, R. G. Kula, N. Yoshida, H. Iida, and K.-i. Matsumoto, “Who should review my code? A file location-based code-reviewer recommendation approach for Modern Code Review,” in Proceedings of the 22nd International Conference on Software Analysis, Evolution, and Reengineering (SANER), pp. 141-150, 2015.

\bibitem{b13}A. Ouni, R. G. Kula, and K. Inoue, “Search-Based Peer Reviewers Recommendation in Modern Code Review,” in Proceedings of the 32nd IEEE International Conference on Software Maintenance and Evolution (ICSME), pp. 367-377, 2016.

\bibitem{b14}Z. Li, S. Lu, D. Guo, N. Duan, S. Jannu, G. Jenks, D. Majumder, J. Green, A. Svyatkovskiy, S. Fu, and N. Sundaresan, “Automating code review activities by large-scale pre-training,” in Proceedings of the 30th ACM Joint European Software Engineering Conference and Symposium on the Foundations of Software Engineering (ESEC/FSE), pp. 1035-1047, 2022.

\bibitem{b15}M. B. Zanjani, H. Kagdi, and C. Bird, “Automatically Recommending Peer Reviewers in Modern Code Review,” IEEE Transactions on Software Engineering, vol. 42, no. 6, pp. 530-543, 2016.

\bibitem{b16}T. Hirao, S. McIntosh, A. Ihara, and K. Matsumoto, “The review linkage graph for code review analytics: a recovery approach and empirical study,” in Proceedings of the 27th ACM Joint Meeting on European Software Engineering Conference and Symposium on the Foundations of Software Engineering (ESEC/FSE), pp. 578-589, 2019.

\bibitem{b17}P. Thongtanunam, and A. E. Hassan, “Review Dynamics and Their Impact on Software Quality,” IEEE Transactions on Software Engineering, vol. 47, no. 12, pp. 2698-2712, 2021.

\bibitem{b18}Y. Yu, H. Wang, G. Yin, and T. Wang, “Reviewer recommendation for pull-requests in GitHub: What can we learn from code review and bug assignment?,” Information and Software Technology, vol. 74, pp. 204-218, 2016.

\bibitem{b19}H. Ying, L. Chen, T. Liang, and J. Wu, “EARec:Leveraging Expertise and Authority for Pull-Request Reviewer Recommendation in GitHub,” in Proceedings of the 3rd International Workshop on CrowdSourcing in Software Engineering (CSI-SE), pp. 29-35, 2016.

\bibitem{b20}E. Sülün, E. Tüzün, and U. Doğrusöz, “RSTrace+: Reviewer suggestion using software artifact traceability graphs,” Information and Software Technology, vol. 130, pp. 106455, 2021.

\bibitem{b21}X. Xie, X. Yang, B. Wang, and Q. He, “DevRec: Multi-Relationship Embedded Software Developer Recommendation,” IEEE Transactions on Software Engineering, vol. 48, pp. 4357-4379, 2022.

\bibitem{b22}G. Rong, Y. Zhang, L. Yang, F. Zhang, H. Kuang, and H. Zhang, “Modeling review history for reviewer recommendation:A Hypergraph Approach,” in Proceedings of the 44th International Conference on Software Engineering (ICSE), pp. 1381-1392, 2022.

\bibitem{b23} Pull request reviews in GitHub: https://docs.github.com/en/pull-requests/collaborating-with-pull-requests/reviewing-changes-in-pull-requests/about-pull-request-reviews

\bibitem{b24} J. Krüger, J. Wiemann, W. Fenske, G. Saake, and T. Leich, “Do you remember this source code?,” in Proceedings of the 40th International Conference on Software Engineering (ICSE), pp. 764-775, 2018.

\bibitem{b25} G. Navarro, “A guided tour to approximate string matching,” ACM Computing Surveys, vol. 33, no. 1, pp. 31-88, 2001.

\bibitem{b26}H. A. Çetin, E. Doğan, and E. Tüzün, “A review of code reviewer recommendation studies: Challenges and future directions,” Science of Computer Programming, vol. 208, pp. 102652, 2021.

\bibitem{b27} Replication Package of the Paper: https://github.com/cufeinfor/MIRRec 

\bibitem{b28} E. L. Lehmann, J. P. Romano, and G. Casella, Testing Statistical Hypotheses: Springer, 1986.

\bibitem{b29}C. Hannebauer, M. Patalas, S. Stünkel, and V. Gruhn, “Automatically recommending code reviewers based on their expertise: an empirical comparison,” in Proceedings of the 31st IEEE/ACM International Conference on Automated Software Engineering (ASE), pp. 99-110, 2016.

\bibitem{b30}S. Asthana, R. Kumar, R. Bhagwan, C. Bird, C. Bansal, C. Maddila, S. Mehta, and B. Ashok, “WhoDo: Automating Reviewer Suggestions at Scale,” in Proceedings of the 27th ACM Joint Meeting on European Software Engineering Conference and Symposium on the Foundations of Software Engineering (ESEC/FSE), pp. 937-945, 2019.

\bibitem{b31}M. Fejzer, P. Przymus, and K. Stencel, “Profile based recommendation of code reviewers,” Journal of Intelligent Information Systems, vol. 50, no. 3, pp. 597-619, 2017.

\bibitem{b32}M. M. Rahman, C. K. Roy, and J. A. Collins, “CoRReCT: Code Reviewer Recommendation in GitHub Based on Cross-Project and Technology Experience,” in Proceedings of the 38th International Conference on Software Engineering (ICSE), pp. 222-231, 2016.

\bibitem{b33}S. Rebai, A. Amich, S. Molaei, M. Kessentini, and R. Kazman, “Multi-objective code reviewer recommendations: balancing expertise, availability and collaborations,” Automated Software Engineering, vol. 27, no. 3-4, pp. 301-328, 2020.

\bibitem{b34}W. H. A. Al-Zubaidi, P. Thongtanunam, H. K. Dam, C. Tantithamthavorn, and A. Ghose, “Workload-aware reviewer recommendation using a multi-objective search-based approach,” in Proceedings of the 16th ACM International Conference on Predictive Models and Data Analytics in Software Engineering (PROMISE), pp. 21-30, 2020.

\bibitem{b35}Z. Xia, H. Sun, J. Jiang, X. Wang, and X. Liu, “A Hybrid Approach to Code Reviewer Recommendation with Collaborative Filtering,” in Proceedings of the 6th International Workshop on Software Mining (SoftwareMining), pp. 24-31, 2017.

\bibitem{b36}A. Strand, M. Gunnarson, R. Britto, and M. Usman, “Using a context-aware approach to recommend code reviewers,” in Proceedings of the ACM/IEEE 42nd International Conference on Software Engineering: Software Engineering in Practice (ICSE-SEIP), pp. 1-10, 2020.

\bibitem{b37}A. Chueshev, J. Lawall, R. Bendraou, and T. Ziadi, “Expanding the Number of Reviewers in Open-Source Projects by Recommending Appropriate Developers,” in Proceedings of the 36th IEEE International Conference on Software Maintenance and Evolution (ICSME), pp. 499-510, 2020.

\bibitem{b38}X. Xia, D. Lo, X. Wang, and X. Yang, “Who Should Review This Change?: Putting Text and File Location Analyses Together for More Accurate Recommendations,” in Proceedings of the 31st IEEE International Conference on Software Maintenance and Evolution (ICSME), pp. 261-270, 2015.

\bibitem{b39}X. Ye, Y. Zheng, W. Aljedaani, and M. W. Mkaouer, “Recommending pull request reviewers based on code changes,” Soft Computing, vol. 25, no. 7, pp. 5619-5632, 2021.

\bibitem{b46}R. Li, P. Liang, P. Avgeriou, "Code Reviewer Recommendation for Architecture Violations: An Exploratory Study," in Proceedings of the 27th International Conference on Evaluation and Assessment in Software Engineering (EASE), pp. 42-51, 2023.

\bibitem{b40}J. Jiang, Y. Yang, J. He, X. Blanc, and L. Zhang, “Who should comment on this pull request? Analyzing attributes for more accurate commenter recommendation in pull-based development,” Information and Software Technology, vol. 84, pp. 48-62, 2017.

\bibitem{b41}J. Jiang, J.-H. He, and X.-Y. Chen, “CoreDevRec: Automatic Core Member Recommendation for Contribution Evaluation,” Journal of Computer Science and Technology, vol. 30, no. 5, pp. 998-1016, 2015.

\bibitem{b42}Z. Ye, Z. Feng, J. Xiao, Y. Gao, G. Fan, H. Zhang, and S. Chen, "Heterogeneous Graph Neural Network-Based Software Developer Recommendation," in Proceedings of the 18th International Conference on Collaborative Computing: Networking, Applications and Worksharing (CollaborateCom), pp. 433-452, 2022.

\bibitem{b43}X. Xie, B. Wang, and X. Yang, “SoftRec: Multi-Relationship Fused Software Developer Recommendation,” Applied Sciences, vol. 10, no. 12, pp. 4333, 2020.

\bibitem{b44}Z. Liao, Z. Wu, Y. Li, Y. Zhang, X. Fan, and J. Wu, “Core-reviewer recommendation based on Pull Request topic model and collaborator social network,” Soft Computing, vol. 24, no. 8, pp. 5683-5693, 2019.

\bibitem{b45}J. Zhang, C. Maddila, R. Bairi, C. Bird, U. Raizada, A. Agrawal, Y. Jhawar, K. Herzig, and A. van Deursen, “Using Large-scale Heterogeneous Graph Representation Learning for Code Review Recommendations at Microsoft,” in Proceedings of the 45th IEEE/ACM International Conference on Software Engineering: Software Engineering in Practice (ICSE-SEIP), pp. 162-172, 2023.

\balance

\end{thebibliography}
\end{document}